\documentclass[twocolumn]{aastex631}

\received{October 18, 2021}
\revised{November 15, 2021}
\accepted{November 20, 2021}

\shorttitle{Stellar Loci V: Photometric Metallicities of 27 Million FGK Stars based on {\it Gaia} EDR3}
\shortauthors{Xu et al.}

\graphicspath{{./}{my_figures/}}

\begin{document}

\title{Stellar Loci V: Photometric Metallicities of 27 Million FGK Stars based on {\it Gaia} Early Data Release 3}

\author[0000-0003-3535-504X]{Shuai Xu}
\affiliation{Department of Astronomy, Beijing Normal University No.19, Xinjiekouwai St, Haidian District, Beijing, 100875, P.R.China}

\author[0000-0003-2471-2363]{Haibo Yuan}
\affiliation{Department of Astronomy, Beijing Normal University 
No.19, Xinjiekouwai St,
Haidian District, Beijing, 100875, P.R.China}

\author[0000-0002-3651-0681]{Zexi Niu}
\affiliation{National Astronomical Observatories,
Chinese Academy of Sciences,
20A Datun Road, Chaoyang District,
Beijing, China}
\affiliation{University of Chinese Academy of Sciences, 19A Yuquan Road, Shijingshan District, Beijing, China}

\author[0000-0002-9824-0461]{Lin Yang}
\affiliation{
College of Artificial Intelligence, Beijing Normal University No.19, Xinjiekouwai St, Haidian District, Beijing, 100875, P.R.China}

\author[0000-0003-4573-6233]{Timothy C. Beers}
\affiliation{Department of Physics and JINA Center for the Evolution of the Elements
(JINA-CEE), University of Notre Dame, Notre Dame, IN 46556, USA}

\author[0000-0003-3250-2876]{Yang Huang}
\affil{South-Western Institute for Astronomy Research, Yunnan University, Kunming 650500, People’s Republic of China}

\correspondingauthor{Haibo Yuan}
\email{yuanhb@bnu.edu.cn}

\begin{abstract}

We combine LAMOST DR7 spectroscopic data and {\it Gaia} EDR3 photometric data to construct high-quality giant (0.7 $< (BP-RP) <$ 1.4) and dwarf (0.5 $< (BP-RP) < $ 1.5) samples in the high Galactic latitude region, with precise corrections for magnitude-dependent systematic errors in the {\it Gaia} photometry and careful reddening corrections using empirically determined color- and reddening-dependent coefficients.  We use the two samples to build metallicity-dependent stellar loci of {\it Gaia} colors  for giants and dwarfs, respectively. For a given $(BP-RP)$ color, a one dex change in [Fe/H] results in about a 5 mmag change in $(BP-G)$ color for solar-type stars. These relations are used to determine  metallicity estimates from EDR3 colors. Despite the weak sensitivity, the exquisite data quality of these colors enables a typical precision of about $\delta$\,[Fe/H] = 0.2 dex. Our method is valid for FGK stars with $G \leq 16$, [Fe/H] $\geq -2.5$, and $E(B-V) \leq 0.5$.   Stars with fainter $G$ magnitudes, lower metallicities, or larger reddening suffer from higher metallicity uncertainties. With the enormous data volume of {\it Gaia}, we have measured metallicity estimates for about 27 million stars with 10 $< G \leq 16$ across almost the entire sky, including over 6 million giants and 20 million dwarfs, which can be used for a number of studies.  These include investigations of Galactic formation and evolution, the identification of candidate stars for subsequent high-resolution spectroscopic follow-up, the identification of wide binaries, and to obtain metallicity estimates of stars for asteroseismology and exoplanet research.

\end{abstract}

\keywords{stars: fundamental parameters     stars:abundances   methods: data analysis   Galaxy: stellar content}

\section{Introduction} \label{sec:intro} 

The abundance of metals (metallicity; often parametrized by [Fe/H]) is one of the most important stellar parameters, as the atmospheric metallicities of long-lived low-mass stars retain a fossil record of the chemical composition of the interstellar medium at the time and place of their formation.
Thus, a census of stellar metallicity plays an important role in understanding the formation and evolution of the Galaxy.(e.g., \citealt{beers2005,Casagrand2011,Peng2013,Wright2021}). 

Up till now, stellar metallicities have been measured primarily from spectroscopic surveys, such as the Sloan Digital Sky Survey (SDSS; \citealt{SDSS2000,Yanny2009}), the Radial Velocity Experiment (RAVE; \citealt{RAVE2006}), the Large Sky Area Multi-Object Fiber Spectroscopic Telescope (LAMOST; \citealt{LAMOSTDR5}), the Apache Point Observatory Galactic Evolution Experiment (APOGEE; \citealt{Majewski2016}), and the GALactic Archaeology with HERMES (GALAH; \citealt{GALAH2018}) survey. This approach has advantages and disadvantages.  It can, for example, achieve an accuracy of 0.03 dex -- 0.1 dex for individual elements from spectra with high signal-to-noise ratios (S/N $>50$) (e.g., \citealt{Garcia2016,Xiang2019,SDSSfeh,RAVEfeh,GALAHfeh}). However, spectroscopic observations and analysis are time-consuming compared to photometric studies. 
Spectroscopic surveys also potentially suffer from strong selection effects, compromising interpretations of the resulting metallicity distributions.

Stellar metallicities can be also obtained through an analysis of photometric data (e.g., \citealt{Ivezic2008,Starkenburg2017,Casagrande2019,Huang2019,huang2021,Whitten2019,Whitten2021,an2020, an2021a,an2021b,bonifacio2021}). Although stellar colors are mostly dependent on effective temperature, they are also influenced by the atmospheric composition, though to a much lesser extent.
The metallicity dependence is stronger for bluer colors, because of the presence of a large number of metallic absorption lines in the blue wavelength range.  

Using the re-calibrated SDSS/Stripe 82 photometry \citep{yuan2015recalibrated}, \cite{yuan2015metal} investigated the metallicity dependence and intrinsic widths of the SDSS stellar loci. It was found that a 1 dex decrease in metallicity typically resulted in a 0.20/0.02 mag decrease in the $u - g/g - r$ colors and a 0.02/0.02 mag increase in the $r - i/i - z$ colors, respectively. The intrinsic widths of the SDSS stellar loci are at maximum a few mmag. By simultaneously fitting the intrinsic colors $u - g$, $g - r$, $r - i$, and $i - z$ from SDSS with those predicted by the metallicity-dependent stellar loci, \citet{yuan2015FGK} measured metallicity estimates for a half million FGK stars in SDSS Stripe 82, with a typical precision of 0.1 dex.
Very recently, \cite{zhang2021redgiant} presented metallicity-dependent SDSS stellar color loci for red giant stars. Systematic differences between the metallicity-dependent stellar loci of red giants and main-sequence  stars were found.  They applied the same technique to measure metallicities for red giant stars in Stripe 82, and achieved a typical precision of 0.2 dex.  

Gaia Early Data Release 3 (EDR3; \citealt{GaiaEDR32021}) has provided the best photometric data to date, obtaining colors of unprecedented mmag precision for more than one billion stars. Such a huge data volume and high data quality make it an ideal database to estimate photometric metallicities for enormous numbers of stars. Indeed, \citet{huang2021} have combined narrow-band photometry from the SkyMapper Southern Survey (SMSS; \citealt{wolf2018,onken2019}) with broad-band {\it Gaia} colors to obtain metallicity estimates for some 24 Million giant and dwarf stars in the Southern Hemisphere (limited by the footprint of SMSS), and demonstrated that excellent results ($\delta$\,[Fe/H] $\sim 0.20-0.25$\,dex) can be obtained for stars with metallicities as low as [Fe/H] $\sim-3.5$, due to the additional sensitivity to metallicity from use of the SMSS $u$- and $v$-band filters along with the $Gaia$ data. \citet{bonifacio2021} employed a similar approach, combining broad-band SDSS photometry with $\it Gaia$ photometry and parallaxes to obtain estimates of metallicity for around 24 Million main-sequence turnoff stars, achieving somewhat lower precision, as expected. 

In this work, we construct the metallicity-dependent stellar loci for {\it Gaia} colors {\it alone}, and investigate the photometric metallicity precision that can be estimated from these very broad-band colors.
The paper is organized as follows. In Section \ref{sec:data}, we describe our data and the methods employed.
The results are presented and tested in Section \ref{sec:result}. We apply this approach to the entire {\it Gaia} EDR3 data (subject to a number of color and latitude cuts), and obtain metallicities for over 27 million stars covering almost the entire sky in Section\,\ref{sec:finalsample}. Section \ref{sec:con} presents a summary as well as future perspectives.

\section{Data and Methodology} \label{sec:data}

In order to construct our training and test samples, we cross-match {\it Gaia} EDR3 and LAMOST Data Release 7 (DR7).  LAMOST has accumulated more than 10 million low-resolution spectra (R $\approx$ 1800) of varying quality. Stellar atmospheric parameters for millions of stars 
are derived by its stellar parameter pipeline (LASP; \citealt{wu2011,LAMOSTDR5}), with a typical precision of 0.1 dex for spectra of sufficient quality and signal-to-noise.  T. Beers (private communication) has refined estimates of the stellar metallicities for LAMOST stars with [Fe/H] $\leq -1.8$ by application of the non-SEGUE Stellar Parameter Pipeline (n-SSPP; \citealt{Beers2014,Beers2017}), which is based on a subset of the methods originally developed for the SEGUE Stellar Parameter Pipeline (SSPP; \citealt{Lee2008a,Lee2008b,Lee2008c,Lee2011,Lee2013}).

For {\it Gaia} photometry, despite its mmag-level precision, magnitude-dependent systematic errors in the magnitudes and colors have been found
in its DR2 (e.g., \citealt{ma2018,weiler2018,niu2021DR2}) and EDR3.
Using the spectroscopic information for about 0.7 million stars from LAMOST DR7, \cite{Niu2021EDR3} have provided precise corrections to the Gaia EDR3 colors with a spectroscopy-based stellar color regression method, achieving an internal calibration precision of about 1 mmag for the colors $G-RP$ and $BP-RP$, respectively. These corrections are applied in this work. Note the correction of \cite{Niu2021EDR3} has taken into account the $G$ magnitude corrections recommended by the {\it Gaia} team \citet{Riello2021}. To match with the magnitude range of \cite{Niu2021EDR3}, the $G$ magnitude of our data is restricted to between 9.5 and 17.5. For {\it Gaia} parallaxes, the official correction is applied to the reported EDR3 values, as validated by \citet{huangoffest}.

\subsection{Reddening Corrections}

\begin{figure}[htbp]
\centering
\includegraphics[width=1.0\linewidth]
{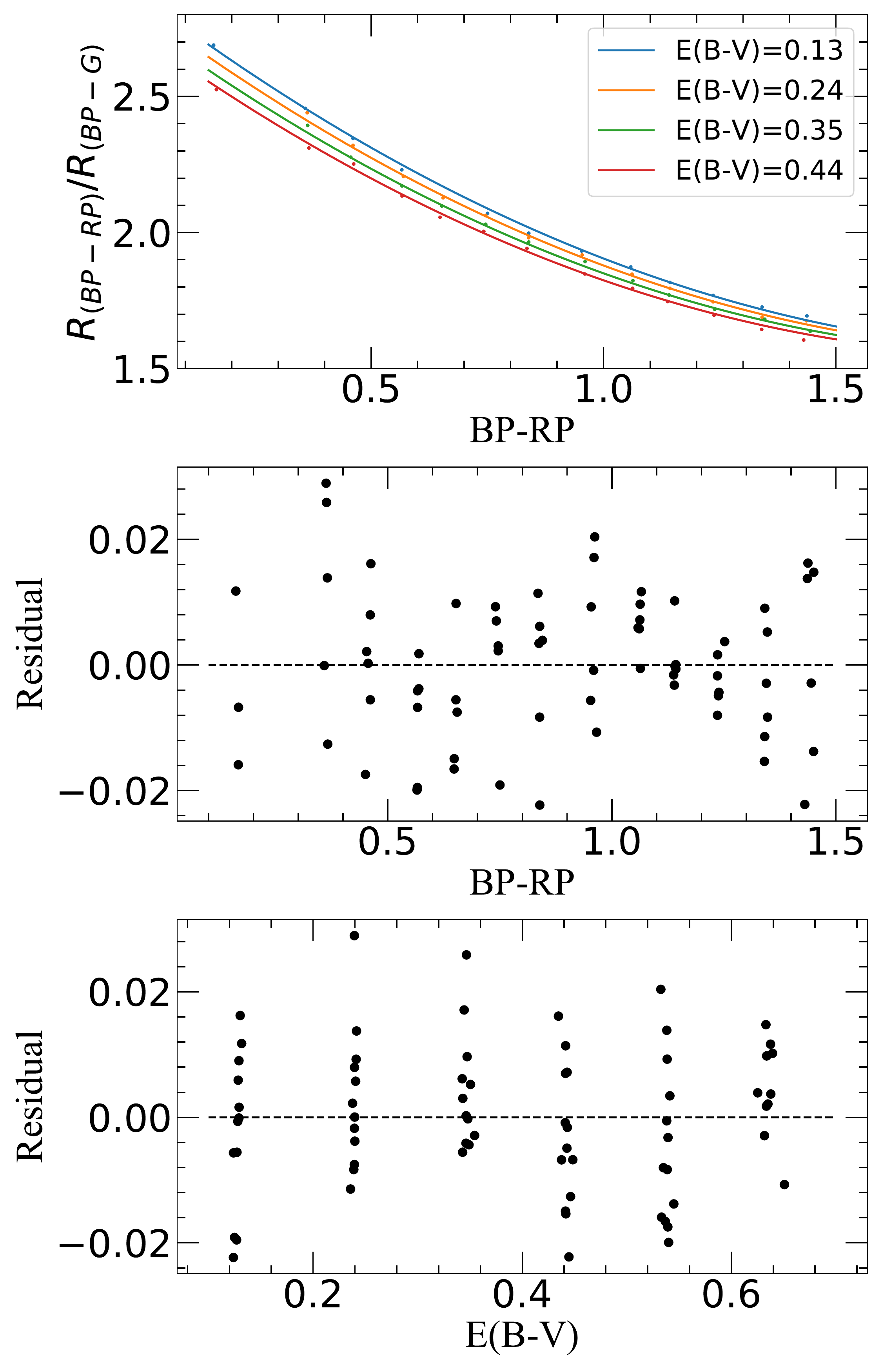}
\caption{
Top panel: $ R_{\rm BP-RP}/R_{\rm BP-G}$, as a function of $(BP-RP)$, for different reddening values. The dots are empirically determined values; the lines are the fitting results. 
Middle and bottom panels: 
Fitting residuals, as a function of $(BP-RP)$ and $E(B-V)$, respectively. In these panels a dashed line is provided at the 0.00 residual level for reference.}
\label{fig:R}
\end{figure}

Due to the very broad passbands of {\it Gaia}, reddening corrections must be carefully performed.
The \citeauthor*{SFD1998} (\citeyear{SFD1998}, hereafter SFD) dust reddening map is used, along with empirical color- and reddening-dependent coefficients for $(BP-G)$ and $(BP-RP)$.
\cite{Niu2021EDR3} have estimated empirical temperature- and reddening-dependent reddening coefficients of $(BP-G)$ and $(BP-RP)$ with respect to the SFD map. These coefficients work well in most cases, but they require knowledge of effective temperature and are fitted independently.
We also note that the errors of our method are more sensitive to their ratios. 
Therefore, we first repeat the work of \cite{Niu2021EDR3} to estimate $R_{\rm BP-RP}$, but as a function of $(BP-RP)_0$ and $E(B-V)$. Then we fit $R_{\rm BP-G}/R_{\rm BP-RP}$ to obtain $R_{\rm BP-G}$. Here stellar reddening values of $E(BP-RP)$ and $E(BP-G)$  are derived 
using the star-pair technique (e.g., \citealt{yuan2013}, \citealt{Ruoyi2020}).
Note that the SFD map over-estimates reddening by about 14 per cent (e.g., 
\citeauthor{Arce1999}, \citealt{Schlafly2011}, \citealt{yuan2013}); 
these systematic errors in the SFD map are naturally taken into account by our empirical coefficients.
The results are as follows:

\begin{eqnarray}
R_{\rm BP-RP} = &1.6876 - 0.5327   X - 1.0268   Y +\nonumber \\
&0.2238   X^2 + 0.0933   X   Y + 0.7177   Y^2 
\end{eqnarray}

\begin{eqnarray}
R_{\rm BP-RP}/R_{\rm BP-G}  = &2.9302 - 1.3140   X - 0.3999   Y + \nonumber \\
&0.3150   X^2 + 0.2114   X   Y -   \nonumber \\
&0.1223   Y^2, 
\end{eqnarray}

\noindent where $X$ is $(BP-RP)_0$, and $Y$ is $E(B-V)$ from the SFD map. 
The fit result for $R_{\rm BP-RP}/R_{\rm BP-G}$ is shown in the top panel of Figure \ref{fig:R}. 
The fitting residuals, as a function of $(BP-RP)$ and $E(B-V)$, are shown in the middle and bottom panels of Figure \ref{fig:R}, respectively. The trends of our coefficients with color are consistent with the color-dependent coefficients of \citet{Casagrande2021}.
Note that, in the above reddening coefficients, $(BP-RP)$ refers to intrinsic colors; 
iterations are needed when using the coefficients.
All colors referred to hereafter are the intrinsic (de-reddened) colors unless otherwise noted.
For the $G$-band extinction coefficient, a constant value of 2.50 (\citealt{chen2019}) is simply adopted, as extinction-corrected $G$ magnitudes are only used 
for classifications of giants and dwarfs in this work.

\begin{figure}[htbp]
\plotone{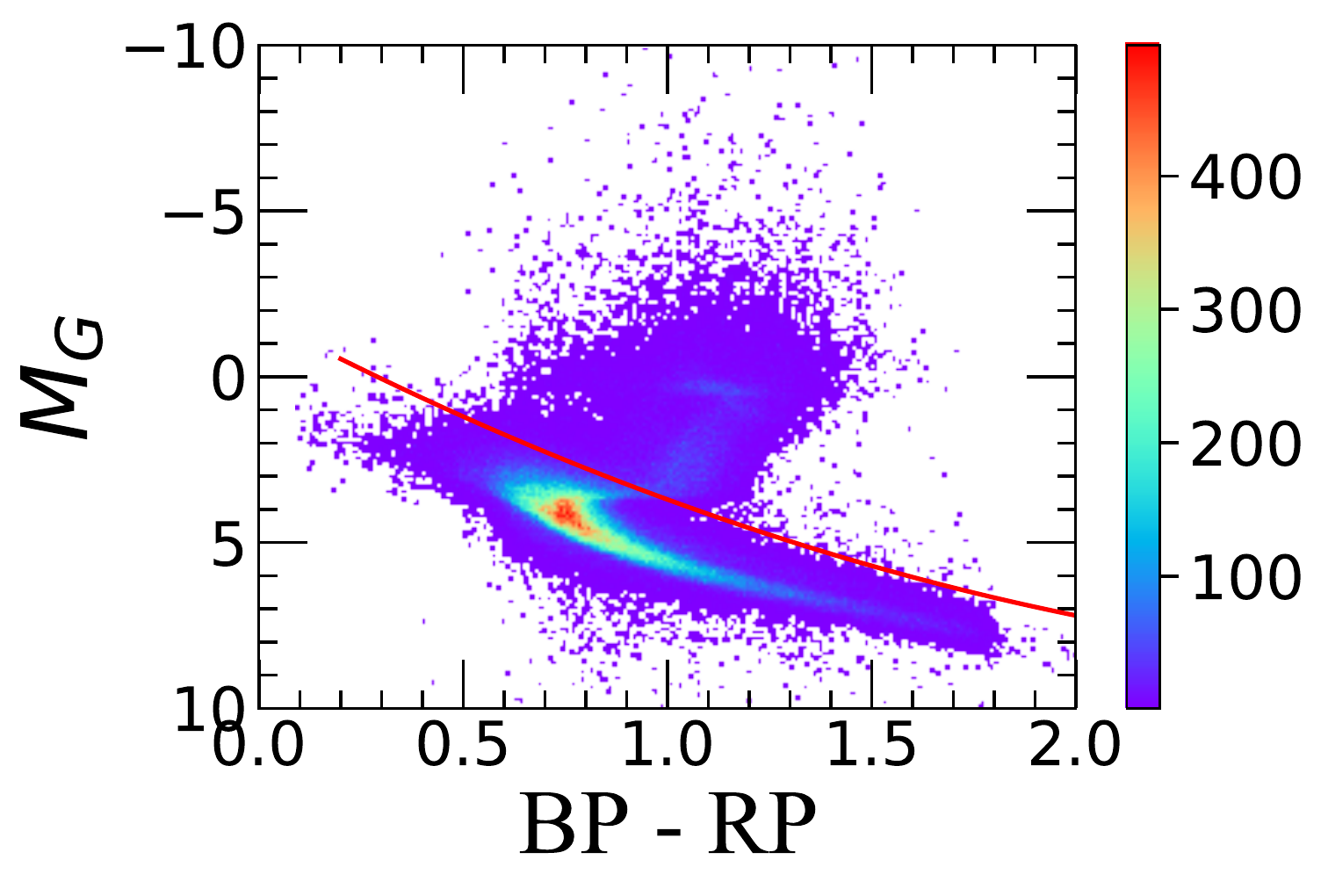}
\caption{Color-magnitude diagram for the training sample. Colors indicate number densities, as shown in the color bar at right. Dwarfs and giants are separated by the red line: $M_G = -{(BP-RP)}^2 + 6.5\times (BP-RP) - 1.8$.}
\label{fig:HRD}
\end{figure}

\subsection{Training Samples}

We divide the sample into dwarfs and giants by an empirical cut, as shown by the red line in Figure \ref{fig:HRD}. The color range is set to be $0.7 < (BP-RP) < 1.5$ for the giants and $0.5 < (BP-RP) < 1.5$ for the dwarfs. 

First, we choose the LAMOST stellar spectra with signal-to-noise ratios in the $g$ band ($S/N_g$) greater than 20 in order to obtain more reliable spectroscopic information. A cut of $|b| > 20^\circ$ is used to avoid high extinction in the low Galactic latitude region. 
We further exclude sources whose $BP$ errors are larger than the median of $BP$ errors for a given magnitude bin. Stars having $G$ magnitudes from 12 to 14 have the best quality \citep{GaiaEDR32021}. This cut is applied for most stars in our training sample. However, for dwarf stars with lower metallicity ([Fe/H] $<-0.8$), we extend the faint limit to $G = 16$ to include a sufficient number of metal-poor dwarf stars.

\begin{figure}[htbp]
\plotone{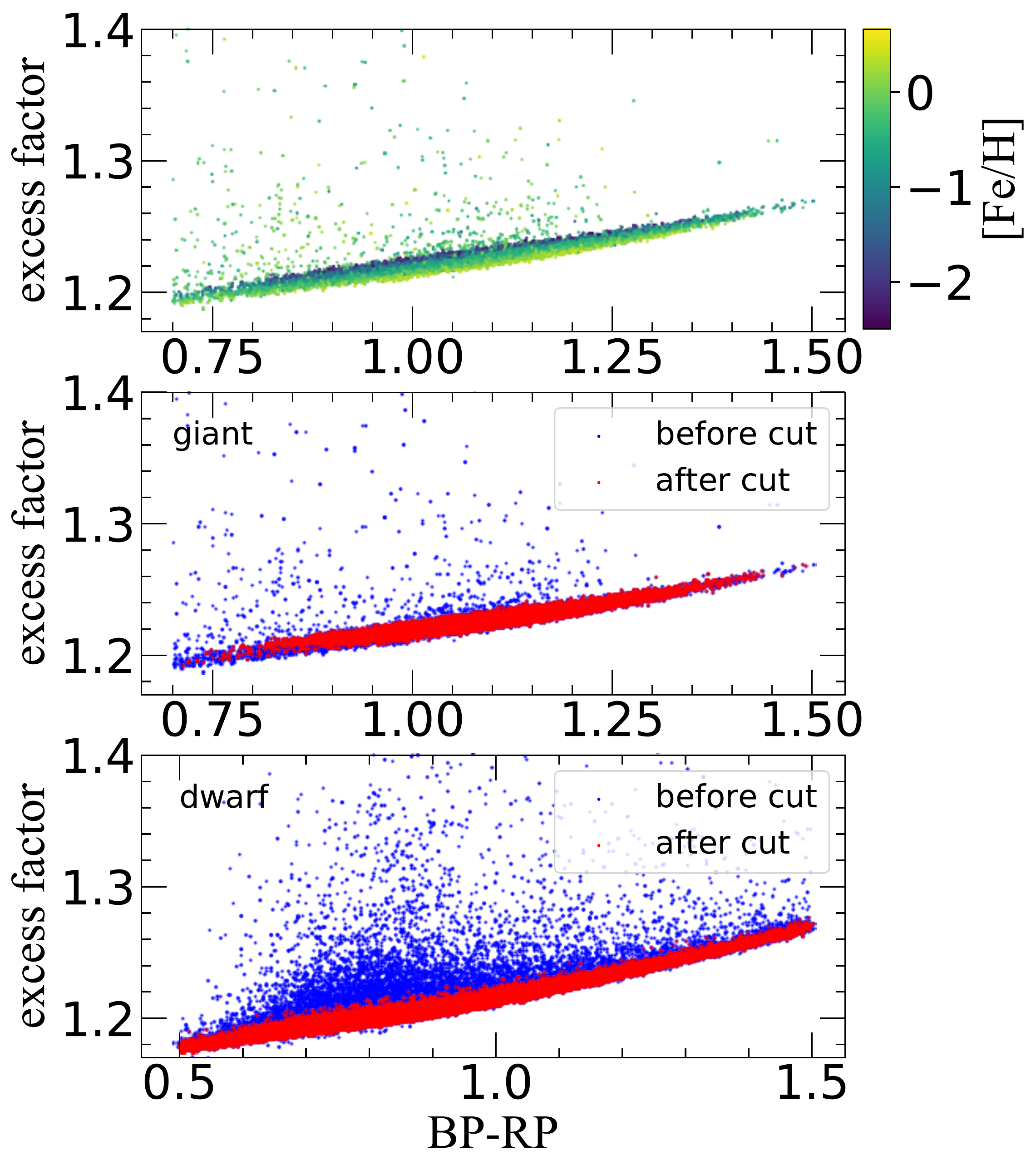}
\caption{The phot\textunderscore bp\textunderscore rp\textunderscore excess\textunderscore factor as a function of $(BP-RP)$. Colors in the top panel indicates [Fe/H] values, as shown in the color bar at right. The middle and bottom panels show before (blue dots) and after (red dots) comparisons as the result of the 3$\sigma$-clipping for giants and dwarfs, respectively.}
\label{fig:excess}
\end{figure}

\begin{figure*}[htbp]
\centering
\includegraphics[width=14cm]{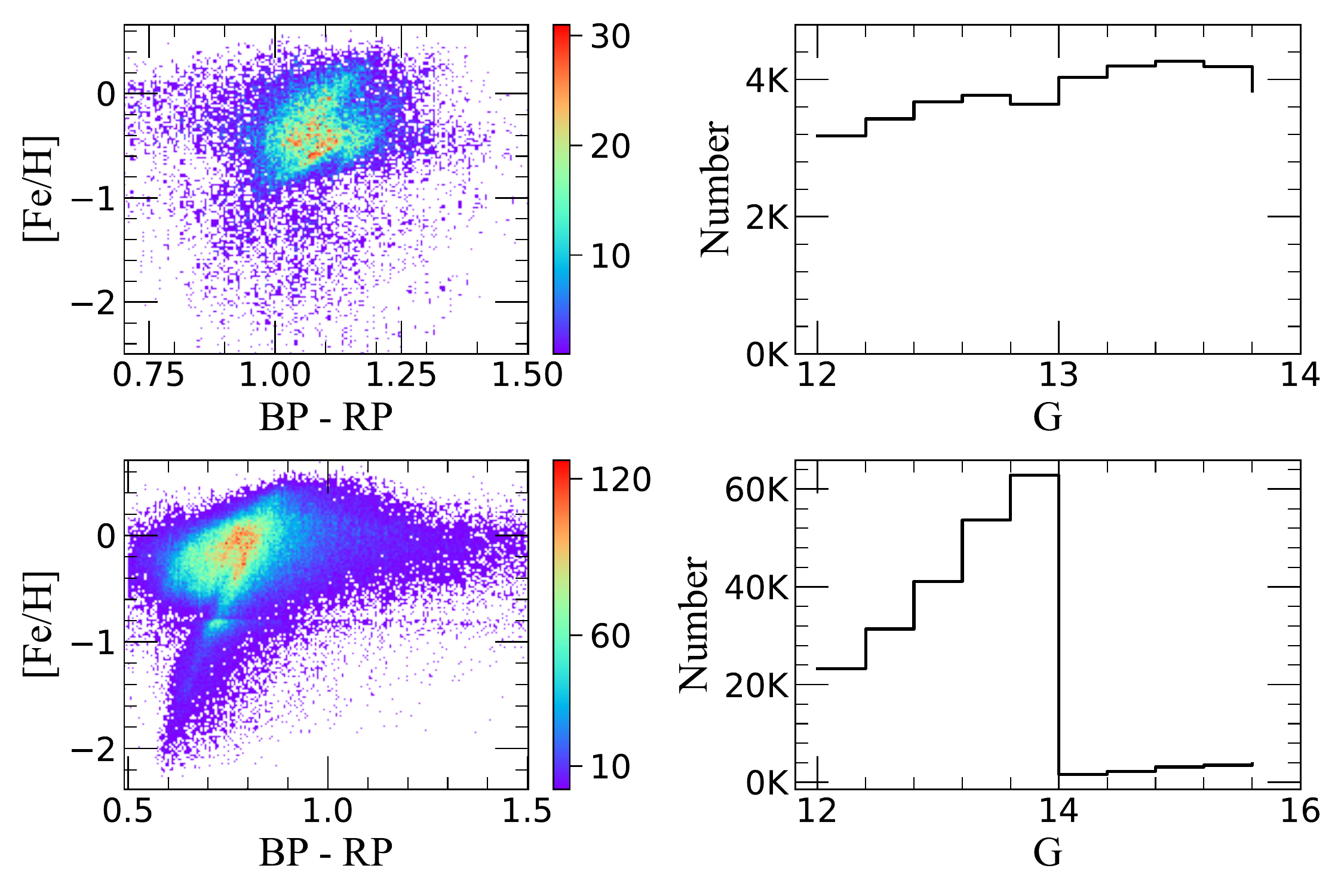}
\caption{Left panels: Distributions of the giant (top) and dwarf (bottom) training samples in the [Fe/H] vs. $(BP - RP)$  plane. The colors indicate the number densities, as shown in the color bar at right. Right panels: $G$ magnitude distributions of the giant (top) and dwarf (bottom) training samples. 
Note that the stars of $14 < G <16$ in the bottom-right panel are metal-poor dwarfs ([Fe/H] $<-0.8$)}. 
\label{fig:train_distrubution}
\end{figure*}

The $BP$ and $RP$ magnitudes may suffer background and contamination issues, since they are obtained from aperture photometry. $Gaia$ provides a phot\textunderscore bp\textunderscore rp\textunderscore excess\textunderscore factor parameter defined as $C = (I_{BP}+I_{RP})/I_G$ to evaluate this situation. As mentioned above, this excess factor suffers magnitude-dependent systematic errors as well. \cite{Yang2021} have made corrections for $Gaia$ magnitudes using approximately 10,000 Landolt standard stars with a machine learning method. The systematic errors in the excess factor are essentially eliminated by applying their results. The excess factor is plotted against $(BP-RP)$ and colored by [Fe/H] in the top panel of Figure \ref{fig:excess}. It can be seen that the excess factor also depends on [Fe/H] for a given $(BP-RP)$.   
A 2nd-order two-dimensional polynomial is  adopted to fit the relation among the excess factor, $(BP-RP)$, and [Fe/H]. We use a Gaussian profile to fit the distribution of residuals and obtain a dispersion, $\sigma$.  Sources that run the risk of lower accuracy are dropped through a 3$\sigma$-clipping process, as shown in Figure \ref{fig:excess}. Finally, we obtain 35,164 giants and 195,155 dwarfs as our training samples. Their distributions in colors and magnitudes are shown in Figure \ref{fig:train_distrubution}.

\subsection{Test Samples}

\begin{figure*}[htbp]
\centering
\includegraphics[width=14cm]{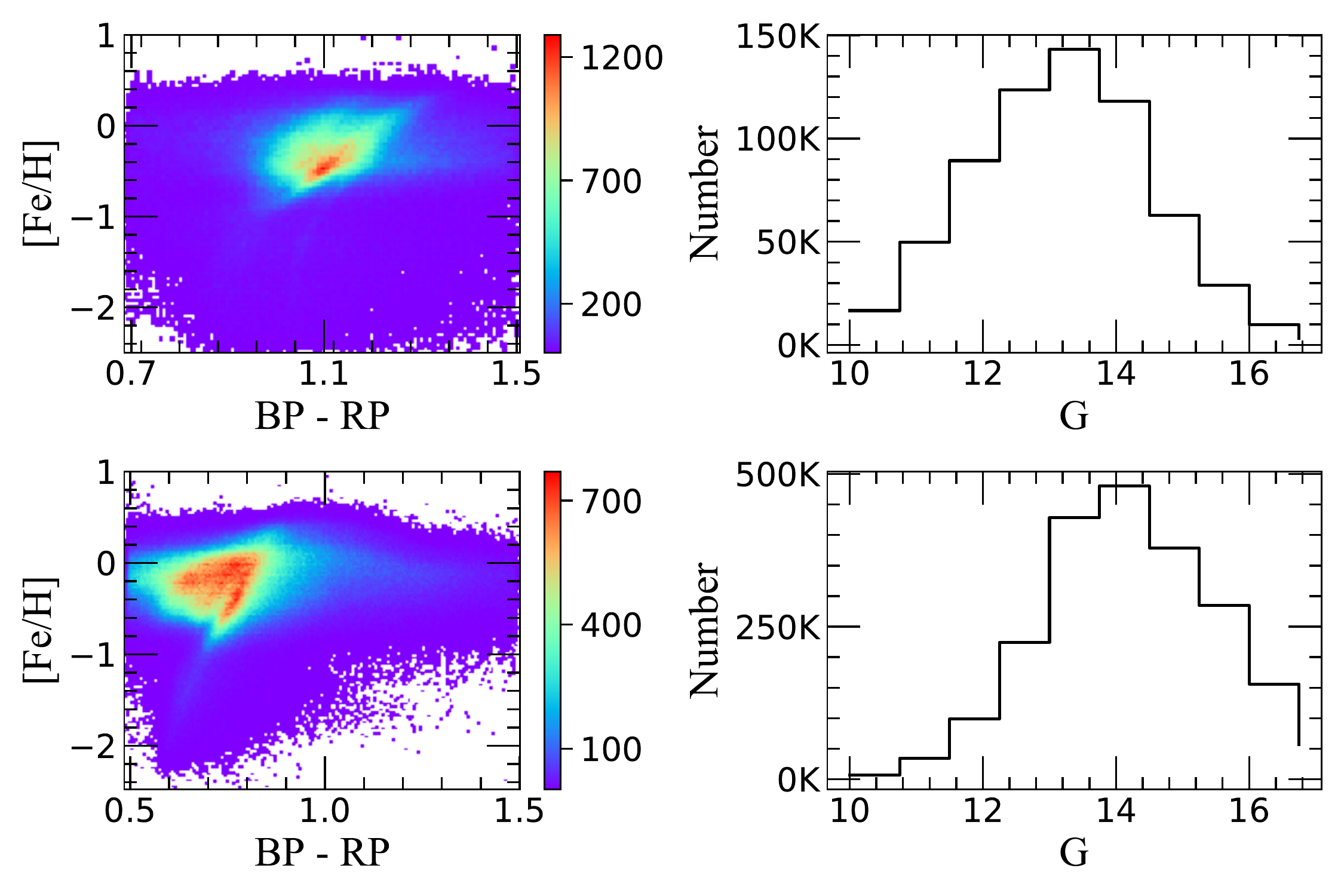}
\caption{Same plot as Figure \ref{fig:train_distrubution}, but for the test samples.}
\label{fig:test_distrubution}
\end{figure*}

As was the case for the training samples, in the assembly of our test samples we require stars with LAMOST spectra having $S/N_g$ greater than 20. Another cut, which requires stars to have distance from the Galactic plane larger than 200 pc is applied, rather than $|b| > 20^\circ$.  In addition, the phot\textunderscore bp\textunderscore rp\textunderscore excess\textunderscore factor is required to be $> 0.09 \times (BP-RP) + 1.15$, in order to excise potentially very inaccurate data. 
Note this criterion is much stricter than the one suggested by the {\it Gaia} 
team (\citealt{Antoja2021}).
The test samples contains 708,268 giants and 2,081,870 dwarfs. Their distributions in colors and magnitudes are shown in Figure \ref{fig:test_distrubution}.
Compared with the training samples, the test samples have larger photometric errors and cover wider magnitude and extinction ranges. 

\subsection{Metallicity-dependent Stellar Loci}

\begin{figure*}[htbp]
\centering
\includegraphics[width=13cm]{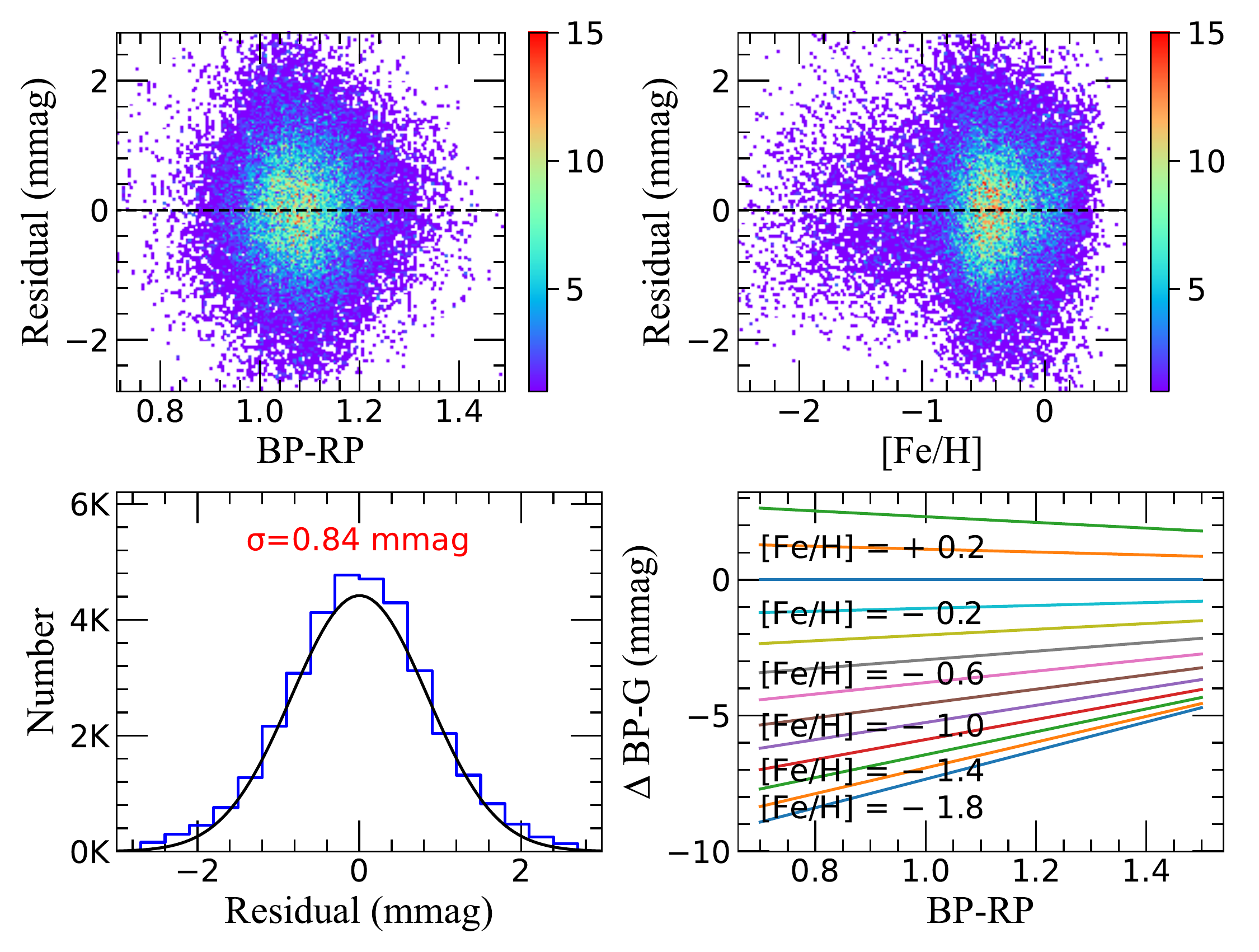}
\caption{Top panels: Fitting residuals, as a function of $(BP-RP)$ color (left) and [Fe/H] (right), for the giant training sample. The colors indicate the number densities, as shown in the color bar at right.   In these panels a dashed line is provided at the 0.00 residual level for reference.  Bottom left: Histogram distributions of the fitting residuals, with the Gaussian fitting profile over-plotted in black.
Bottom right: Variations of stellar loci for different metallicities, relative to the one at [Fe/H] = 0.0. }
\label{fig:train_fit_giant}
\end{figure*}

As shown in \cite{zhang2021redgiant}, metallicity-dependent stellar loci may  differ between giants and dwarfs. 
Therefore, we fit the relations among $(BP-G)$, $(BP-RP)$, and [Fe/H] with a least-squares method separately for the giants and the dwarfs, respectively. 
A 2nd-order two-dimensional polynomial with 6 free parameters is adopted for  giant stars. A 3$\sigma$-clipping procedure is performed to reject outliers when fitting the data.
The relation for giant stars is: 

\begin{eqnarray}\label{eq1}
(BP-G) = &-6.402\times 10^{-3} + 3.352\times 10^{-1} X + \nonumber \\
&8.756\times 10^{-3} Y +8.975\times 10^{-2}   X^2 -\nonumber \\
&3.182\times 10^{-3} XY + 9.230\times 10^{-4}   Y^2,
\end{eqnarray}

\noindent where $X$ is the $(BP-RP)$ color and $Y$ is [Fe/H]. The fitting residuals, as a function of $(BP-RP)$ and [Fe/H], are plotted in the top panels of Figure \ref{fig:train_fit_giant}. We use a Gaussian profile to fit the distribution of residuals; a $\sigma$ of only 0.84 mmag centered at 0.004 mmag is obtained, as shown in the bottom-left panel. The bottom-right panel shows variations of the relation at different metallicities, relative to the one at [Fe/H] = 0.0. It can be seen that a 1 dex change in metallicity results in about a 5 mmag change in $(BP-G)$ color. The sensitivity for metal-rich stars is stronger than that for metal-poor stars, and stronger for bluer stars, as expected. 

\begin{figure*}[htbp]
\centering
\includegraphics[width=13cm]{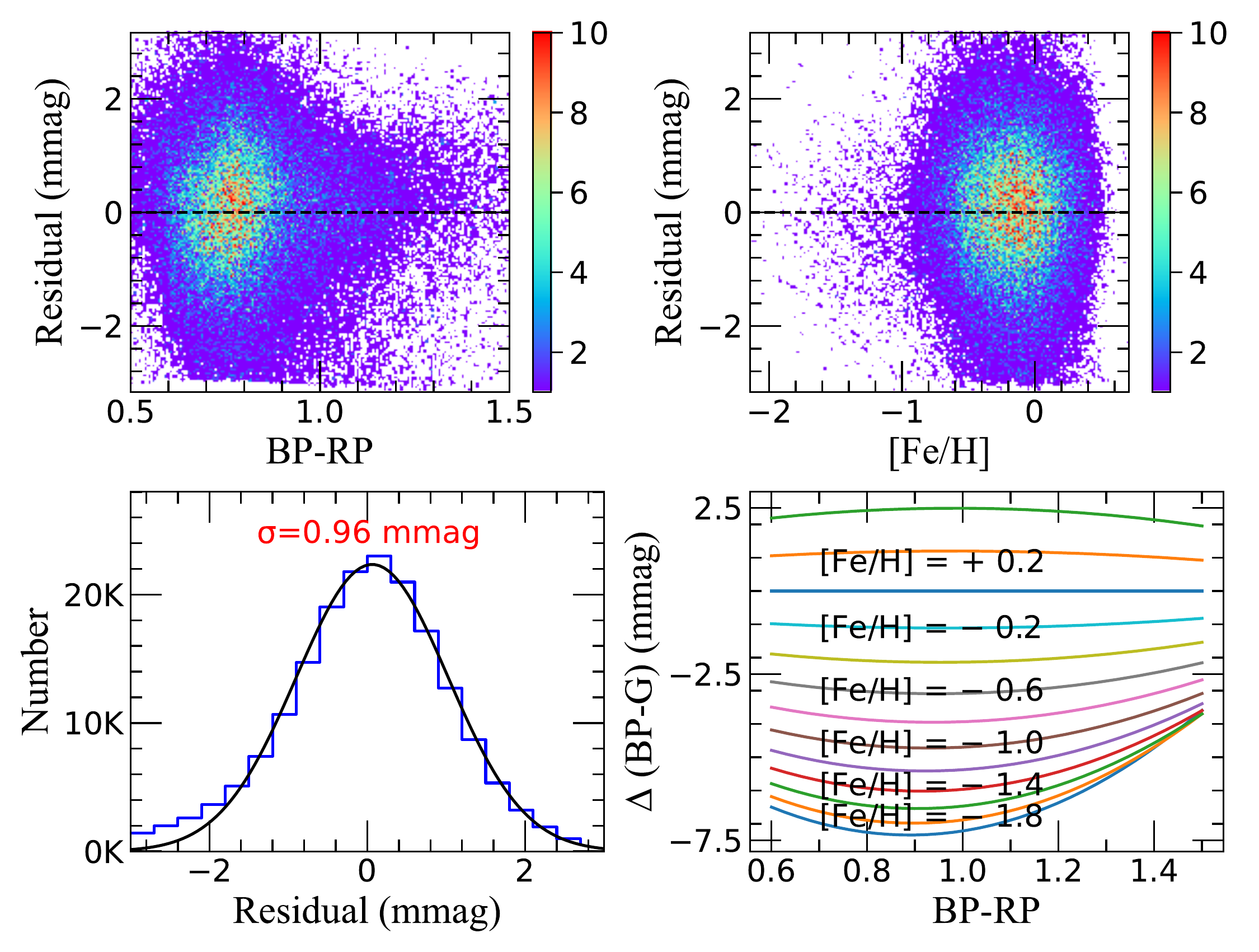}
\caption{Same as Figure \ref{fig:train_fit_giant}, but for the dwarf training sample.}
\label{fig:train_fit_dwarf}
\end{figure*}

Similarly, a 3rd-order two-dimensional polynomial with 10 free parameters is adopted for the metallicity-dependent stellar locus of dwarf stars. The relation is 
\begin{eqnarray}\label{eq2}
(BP-G) = &1.172\times 10^{-2} + 2.762\times 10^{-1} X +\nonumber \\ 
&9.170\times 10^{-4} Y + 1.533\times 10^{-1}  X^2 + \nonumber \\ 
&1.010\times 10^{-2} XY + 6.410\times 10^{-4}   Y^2 - \nonumber \\ 
&2.248\times 10^{-2} X^3 - 5.230\times 10^{-3} X^2Y+\nonumber \\ 
&4.610\times 10^{-4} XY^2 + 0.000 Y^3,
\end{eqnarray}

\noindent where $X$ and $Y$ are the same as in Eqn. \ref{eq1}. The fitting results are shown in Figure \ref{fig:train_fit_dwarf}. The $\sigma$ value is 0.96 mmag, slightly larger than that for giants. 
The tendency of metallicity sensitivity is the same as above, stronger for metal-rich stars than for metal-poor stars, and stronger for F/G/K stars than for A/M stars. 
Note that, due to the presence of unresolved dwarf-dwarf binaries, the distribution of the fitting residuals is asymmetric. Such features have been used previously to study binary fractions of dwarf stars (see, e.g., \citealt{yuan2015binary,niu2021binary}).

\section{Results}\label{sec:result}

In this section, we apply the above relations to estimate photometric metallicities from {\it Gaia} EDR3 colors.
The [Fe/H] calculated by our model is referred to below as  $[\rm Fe/H]_{Gaia}$. We define residuals $\Delta [\rm Fe/H] = [\rm Fe/H]_{Gaia} - [\rm Fe/H]_{\rm LAMOST}$ to evaluate the results. 

\subsection{Results for the Training Samples}

\begin{figure*}[htbp]
\centering
\includegraphics[width=13cm]{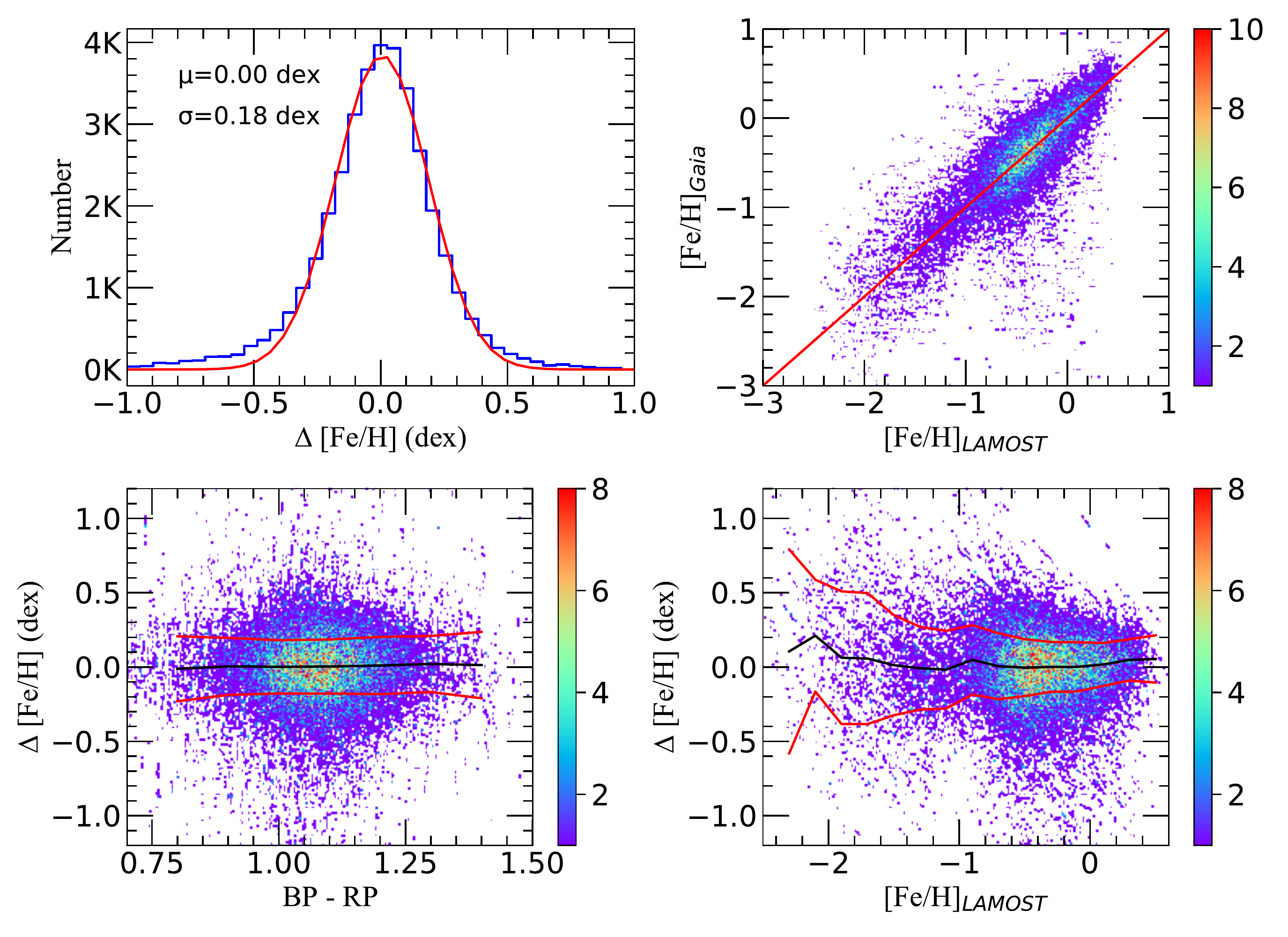}
\caption{Results for the giant training sample. Top left: Histogram of the metallicity differences between those estimated from {\it Gaia} photometry and LAMOST spectroscopy. Gaussian fitting is over-plotted in red; the mean and sigma values are marked.
Top right: {\it Gaia} metallicities vs. LAMOST metallicities. The red line  is the one-to-one line. 
Bottom panels: Metallicity differences, as a function of $(BP-RP)$ color (left) and $[\rm Fe/H]_{LAMOST}$ (right). The black lines indicate the medians; red lines indicate the standard deviations.
}
\label{fig:test_train_giant}
\end{figure*}

\begin{figure*}[htbp]
\centering
\includegraphics[width=13cm]{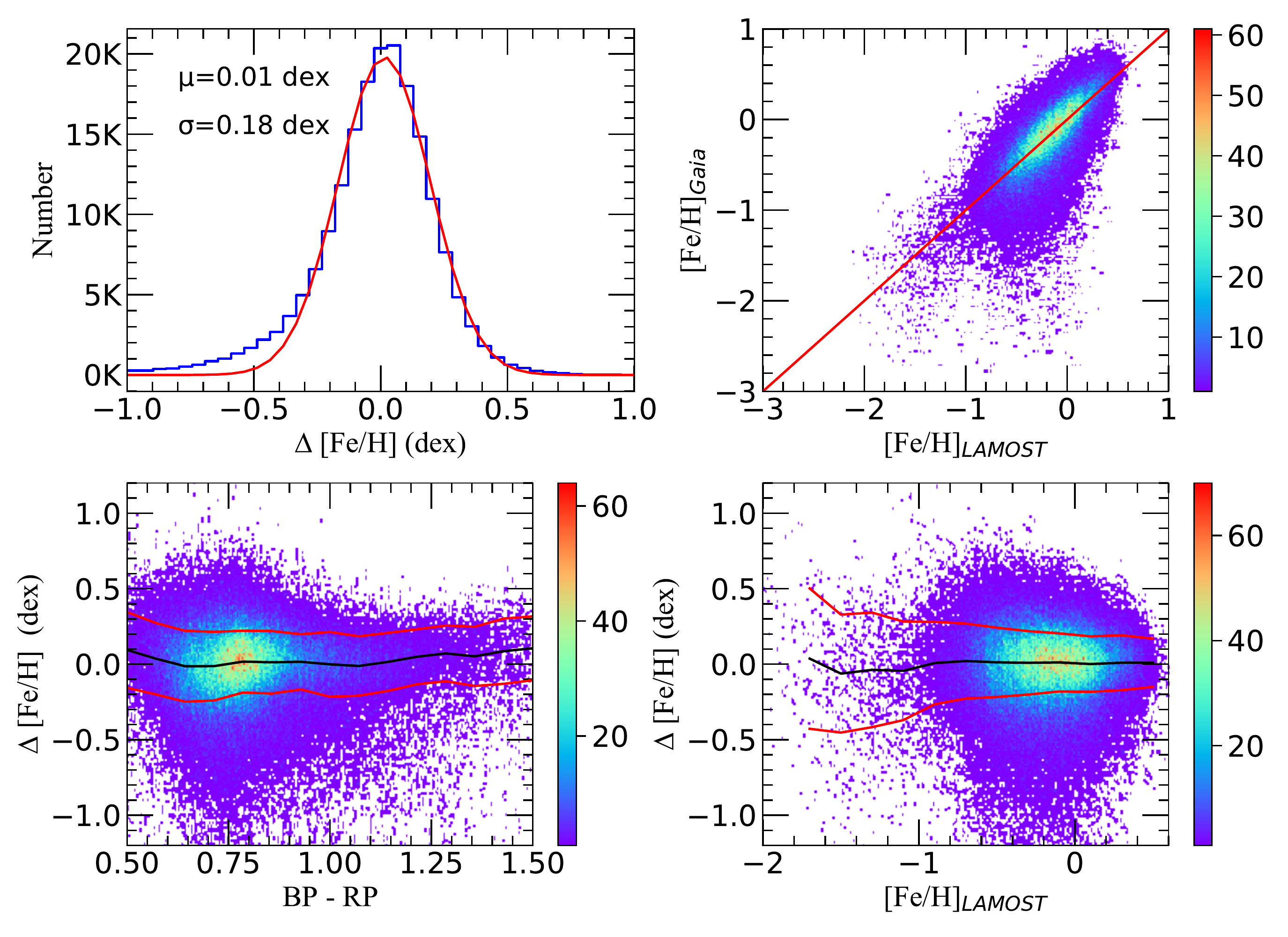}
\caption{Same as Figure \ref{fig:test_train_giant}, but for the dwarf training sample. 
Note that in the top-left panel, the excess on the left side is due to the presence of dwarf-dwarf binary stars.}
\label{fig:test_train_dwarf}
\end{figure*}

We plot the photometric metallicities of the giant and dwarf training samples in Figure \ref{fig:test_train_giant} and Figure \ref{fig:test_train_dwarf}, respectively.
The histogram distributions of $\Delta [\rm Fe/H]$ are displayed in the top-left panels. The top-right panels plot the $[\rm Fe/H]_{Gaia}$ vs. $[\rm Fe/H]_{\rm LAMOST}$. The bottom panels plot $\Delta [\rm Fe/H]$ as a function of $(BP-RP)$ and $[\rm Fe/H]_{LAMOST}$.
Both giants and dwarfs exhibit reasonably good results, with a typical error of 0.18 dex. The median residuals are very close to zero. 
The differences exhibit little variation with $(BP-RP)$, but increase for more metal-poor stars, especially when $[\rm Fe/H]_{\rm LAMOST} < -1.5$, 
as expected.

\subsection{Results for the Test Samples}

\begin{figure*}[htbp]
\centering
\includegraphics[width=16cm]{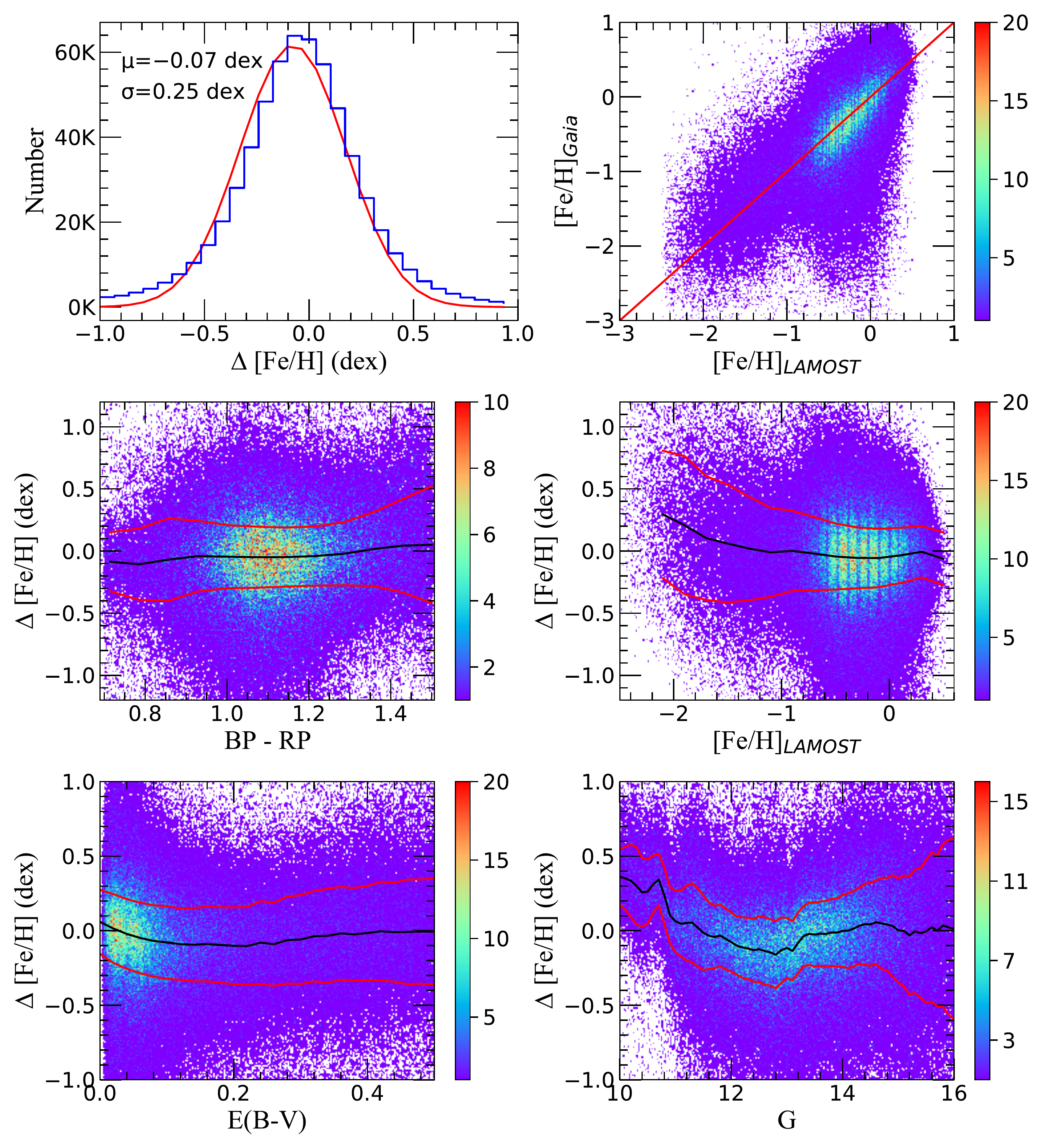}
\caption{Results for the giant test sample.
Top left: Histogram distribution of the metallicity differences that are respectively deduced from {\it Gaia} photometry and LAMOST spectroscopy. Gaussian fitting is over-plotted in red; the mean and sigma values are marked.
Top right: LAMOST metallicities vs. {\it Gaia} metallicities. The red line is the one-to-one line. 
Middle panels: Metallicity differences, as a function of $(BP-RP)$ (left) and [Fe/H] (right). 
Bottom panels: Metallicity differences, as a function of reddening (left) and $G$ magnitude (right).}
\label{fig:test_test_giant}
\end{figure*}

\begin{figure*}[htbp]
\centering
\includegraphics[width=16cm]{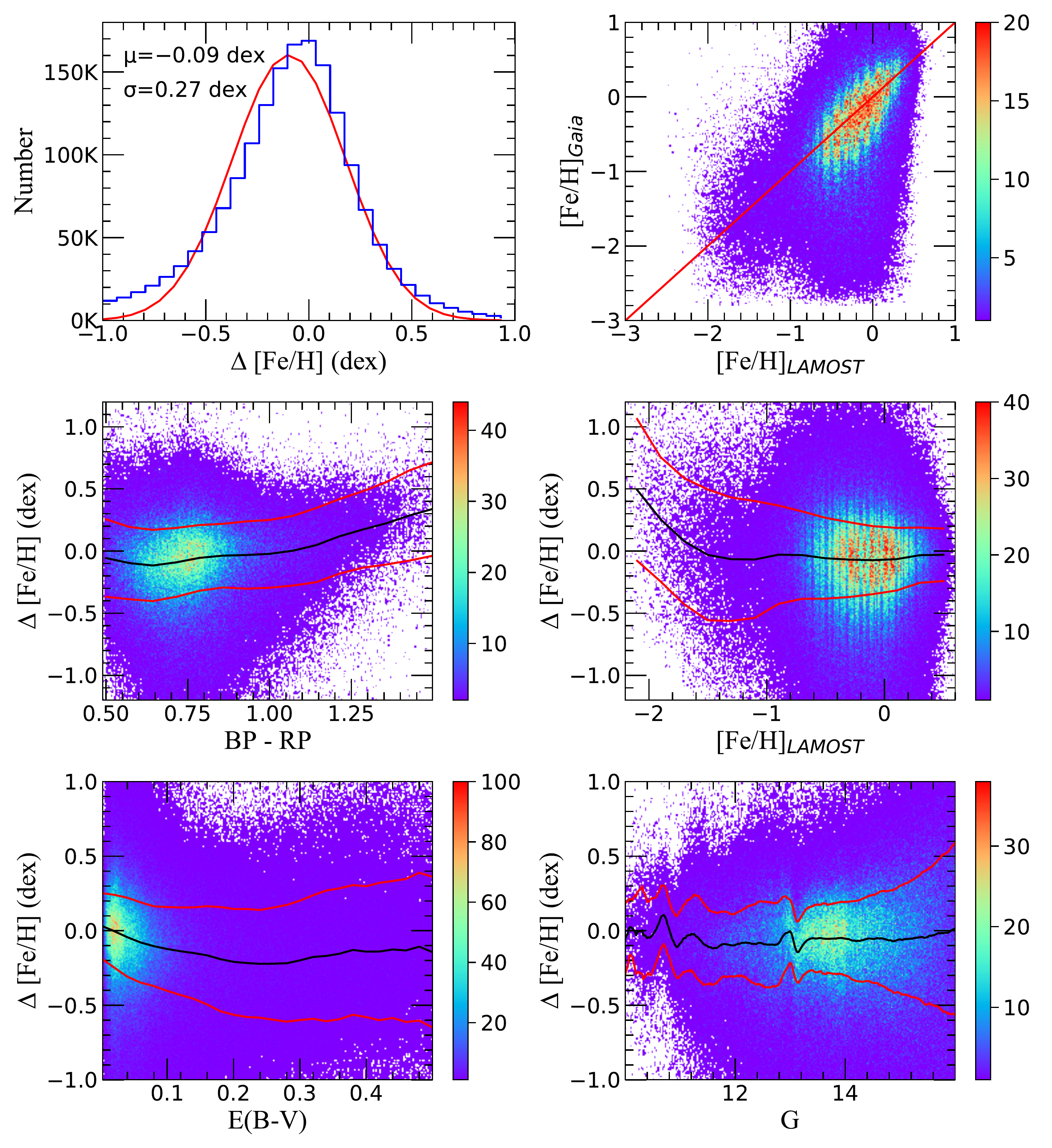}
\caption{Same plot as Figure \ref{fig:test_test_giant}, but for the dwarf test sample. }
\label{fig:test_test_dwarf}
\end{figure*}

\begin{figure}[htbp]
\plotone{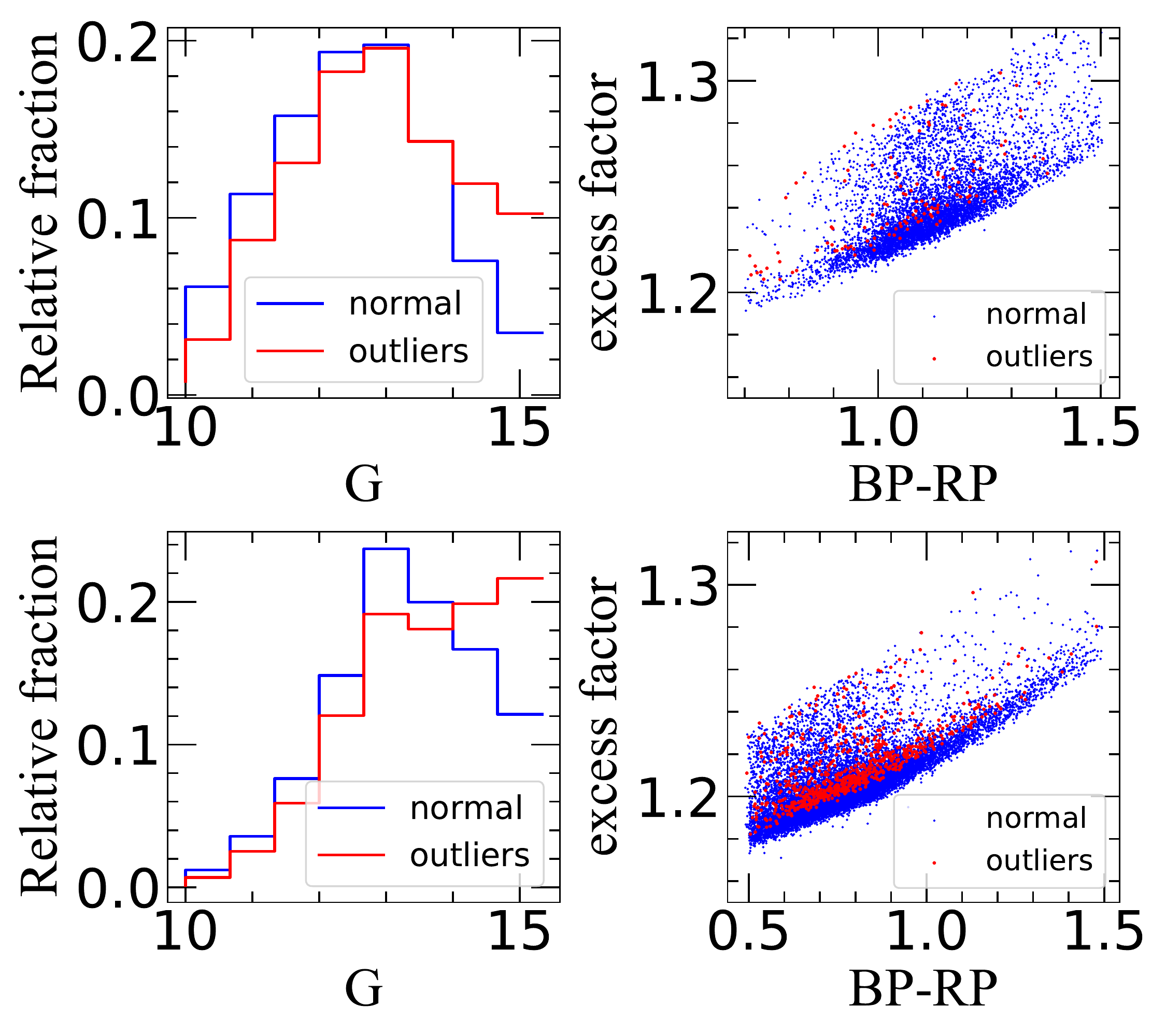}
\caption{Analysis of the outliers. Top left: $G$ magnitude distributions of the normal (blue line) and outliers (red line) for the giants. Top right: The phot\textunderscore bp\textunderscore rp\textunderscore excess\textunderscore factor, as a function of $(BP-RP)$, for the giants. 
Bottom: Same as the top panel, but for the dwarfs. To avoid crowding, only 1 per cent of randomly selected stars are plotted.}
\label{fig:big_err}
\end{figure}

We use the same methods as above in order to obtain the photometric [Fe/H] estimates for the test samples. 
The results are plotted in Figures \ref{fig:test_test_giant} and \ref{fig:test_test_dwarf}. 
As shown in the top-left panels, the standard deviations are 0.25 dex for the giants and 0.27 dex for the dwarfs, somewhat larger than those for the training samples, due primarily to the larger photometric errors in the test samples. 
The mean offsets are negative, $-0.07$\,dex for giants and $-0.09$\,dex for dwarfs, probably due 
to the effect of photometric errors. Since there are more metal rich than metal poor stars, errors will scatter more metal rich stars into the metal poor loci than vice versa.
In the top-right panels, there are some outliers on the bottom right. We define these sources with $[\rm Fe/H]_{Gaia} - [\rm Fe/H]_{\rm LAMOST} < -1$ dex as outliers, making up $2.1\%$ of the giant test sample and $4.1\%$ of the dwarf test sample.
As shown in Figure \ref{fig:big_err}, for the giants, the outliers have a larger phot\textunderscore bp\textunderscore rp\textunderscore excess\textunderscore factor and fainter $G$ magnitudes, so naturally have larger errors. For the dwarfs, 
there are obviously more outliers due to unresolved binaries, which barely affect the giant stars \citep{niu2021binary}, in addition to the two above reasons. 

Because the test samples cover wider ranges of $G$ magnitudes and extinctions comparing with the training samples, we consider their impact on the [Fe/H] residuals. The middle and bottom panels of Figures \ref{fig:test_test_giant} and \ref{fig:test_test_dwarf} display, in turn, $\Delta$[Fe/H] as a function of $(BP-RP)$, [Fe/H]$_{\rm LAMOST}$, $E(B-V)$, and $G$ magnitude. 
For the giants, one can see that the errors hardly change with the color. The errors increase slowly with reddening. 
The errors are larger for more metal-poor 
stars as found for the training sample. However, a systematic deviation gradually happens for [Fe/H]$_{\rm LAMOST} < -2$. This arises because we do not have enough stars with [Fe/H]$_{\rm LAMOST} < -2$ in the training sample, 
as seen in Figure \ref{fig:train_distrubution}. 
A small problem in \cite{Niu2021EDR3} was also found from this analysis. These authors ignored the color dependency of the calibrations for stars of $G < 11.5$ mag 
because it is very weak (less than a few mmag). 
But in this work, even an error of 1 mmag could affect the results. As a consequence, a magnitude-dependent systematic error exists in our model. We provide correction factors for application of our model for giants in the range  $10 < G < 11.5$ in Table \ref{table1}.  This table lists the empirical offsets in metallicity, which are to be subtracted from the predicted [Fe/H] of our model for the brighter stars. 
The results for the dwarfs are similar.  A systematic deviation occurs for stars with [Fe/H]$_{\rm LAMOST} < -1.7$ for similar reasons.
A systematic deviation up to 0.3 dex is also noted for stars with $(BP-RP)>1.2$. 

There are no stars with [Fe/H] $< -2.5$ in the LAMOST DR7 data set due to the limitations of the LASP pipeline.
To further test the limits of our method, we use metallicity estimates for LAMOST stars with [Fe/H] $\leq -1.8$ 
using the n-SSPP, as described above, which are valid down to [Fe/H] about $-4.0$. 
Problematic stars were rejected after visual inspection (by Beers).  Only stars with $G \leq 16$ and spectra with $S/N_g$ $>$ 20 are used.  The test results show that it is very challenging to reach below $-2.5$ with {\it Gaia} EDR3 colors alone.

\setlength{\tabcolsep}{1.5mm}{
\begin{table}[htbp]
\footnotesize
\centering
\caption{$[\rm Fe/H]_{Gaia}$ Correction Factors for Giants with\\ $10 < G < 11.5$.}
\begin{tabular}{cc|cc|cc} 
    
\textbf{G} & \textbf{$\Delta$[Fe/H]} & \textbf{G} & \textbf{$\Delta $ [Fe/H]} & \textbf{G} & \textbf{$\Delta$[Fe/H]} \\
           & (dex)  &       &  (dex)   &       &   (dex)  \\
\hline
10.00 & +0.385 & 10.50 & +0.202 & 11.00 & +0.101  \\
10.05 & +0.296 & 10.55 & +0.289 & 11.05 & +0.080  \\ 
10.10 & +0.406 & 10.60 & +0.330 & 11.10 & +0.063  \\
10.15 & +0.378 & 10.65 & +0.328 & 11.15 & +0.064  \\
10.20 & +0.364 & 10.70 & +0.372 & 11.20 & $-$0.007 \\  
10.25 & +0.321 & 10.75 & +0.301 & 11.25 & +0.048  \\
10.30 & +0.341 & 10.80 & +0.260 & 11.30 & +0.066  \\
10.35 & +0.179 & 10.85 & +0.165 & 11.35 & +0.082  \\
10.40 & +0.254 & 10.90 & +0.094 & 11.40 & +0.028  \\ 
10.45 & +0.273 & 10.95 & +0.097 & 11.45 & +0.038  \\  
\hline
\end{tabular}
\label{table1}
\end{table}}

\subsection{An External Test with Open Clusters}

\begin{figure*}[htbp]
\includegraphics[width=17.5cm]{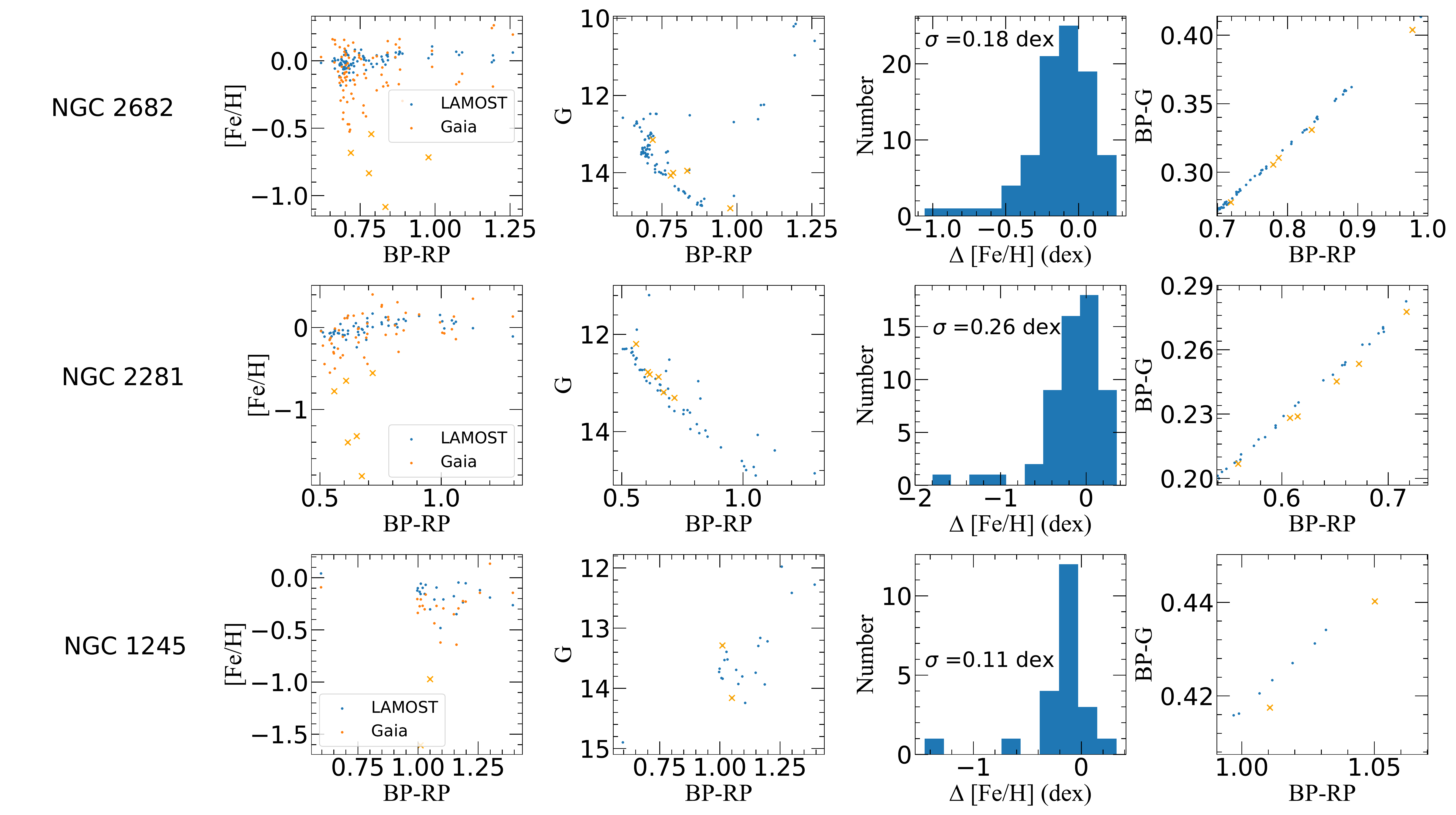}
\caption{Test results for open clusters. The top, middle, and bottom panels are  for NGC\,2682, NGC\,2281, and NGC\,1245, respectively. For each cluster, 
the first column plots [Fe/H] as a function of $(BP-RP)$. The blue and orange dots represent results from LAMOST DR7 and {\it Gaia} EDR3, respectively. The second column is the 
color-magnitude diagram. The third column is the histogram distribution of $\Delta$[Fe/H] between the LAMOST DR7 and {\it Gaia} EDR3 results. The fourth column is the color-color diagram. The orange crosses in the second and fourth columns represent stars with large $\Delta$[Fe/H] values.}
\label{fig:cluster}
\end{figure*}

\cite{cluster} have compiled a list of thousands of known or putative open clusters and their member stars based on {\it Gaia} DR2. 
We cross-match their catalogs with our test samples and obtain 812 sources in 70 open clusters in common. 
NGC\,2682, NGC\,2281, and NGC\,1245 are selected to test our model. 
The results are shown in Figure \ref{fig:cluster}.
We can see that our photometric metallicities are consistent with the LAMOST DR7 results,
with the mean metallicity differences close to zero and the $\sigma$ values in the range 0.11 to 0.26 dex. 
NGC\,1245 has the smallest $\sigma$ value because its member stars are relatively brighter.
There are also a few metallicity outliers, as indicated by orange crosses in the first column of panels. 
Their distributions in the color-magnitude and color-color diagrams suggest that they are binaries or are affected by nearby stars.

\subsection{Loci Differences between Dwarfs and Giants}\label{locus}

\begin{figure}[htbp]
\plotone{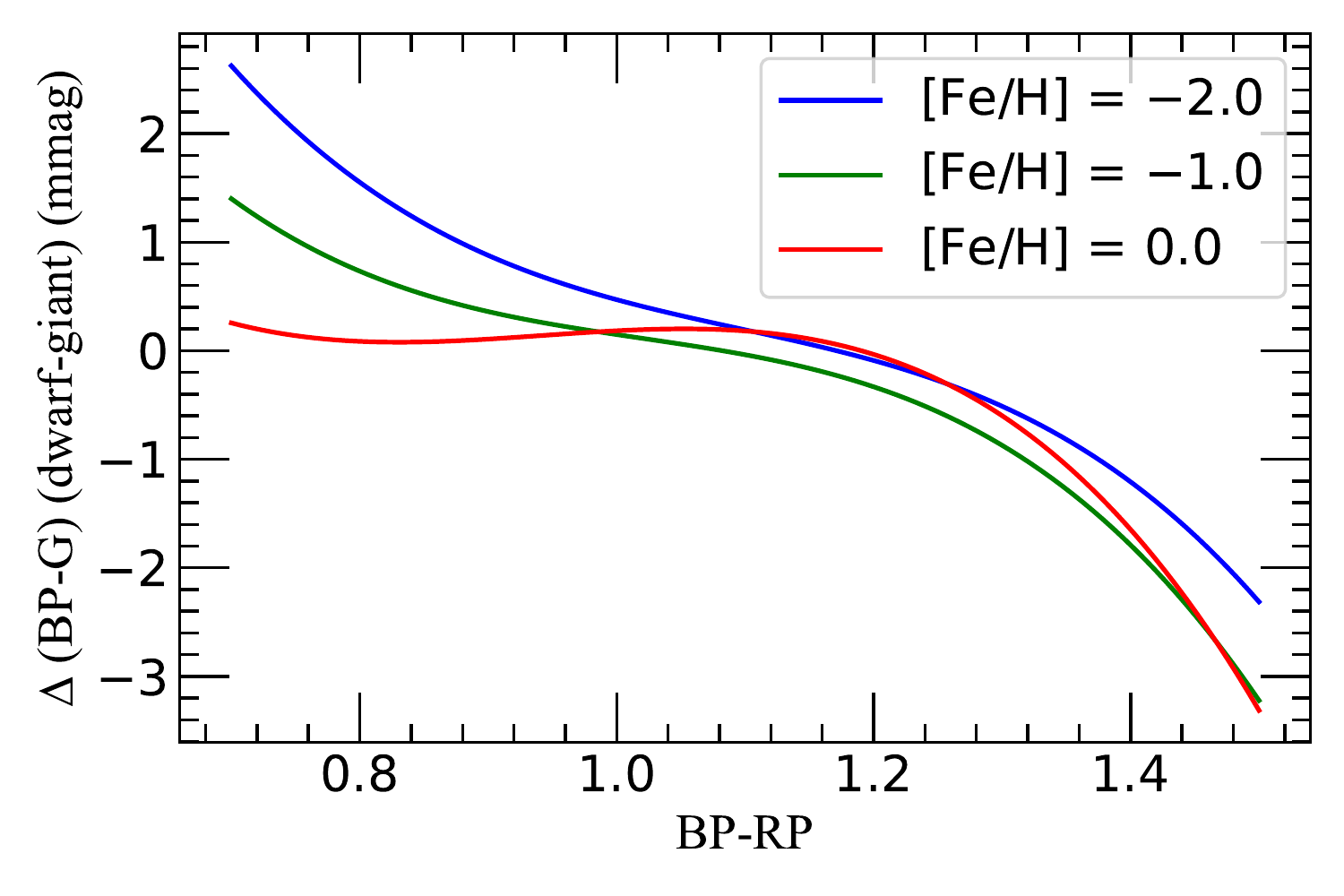}
\caption{
Differences in the loci between dwarfs and giants. The red, green, and blue lines correspond to three different metallicities.}
\label{fig:dwarf_giant_locus}
\end{figure}

As mentioned above, metallicity-dependent stellar loci may be different between dwarfs and giants. 
We define $\Delta (BP-G) = (BP-G)_{\rm dwarf} - (BP-G)_{\rm giant}$ for a given $(BP-RP)$ and [Fe/H]. 
The effect of $(BP-RP)$ and [Fe/H] on $\Delta (BP-G)$ is shown in Figure \ref{fig:dwarf_giant_locus}. 
At [Fe/H] = 0, the difference is almost 0 when $0.7 < BP-RP < 1.2$. 
For stars with $(BP-RP) > 1.0$, the effect on [Fe/H] is small; the difference is almost completely dominated by $(BP-RP)$, 
and becomes larger for redder stars. 
For stars with $(BP-RP) < 1.0$,  the difference is larger at lower metallicities and for bluer stars. 
The largest difference is about 2 mmag, 
suggesting that the largest error of our model could 
reach about 0.4 dex if a source is mistakenly classified.

\section{The Final Sample}\label{sec:finalsample}

\begin{figure}[htbp]
    \plotone{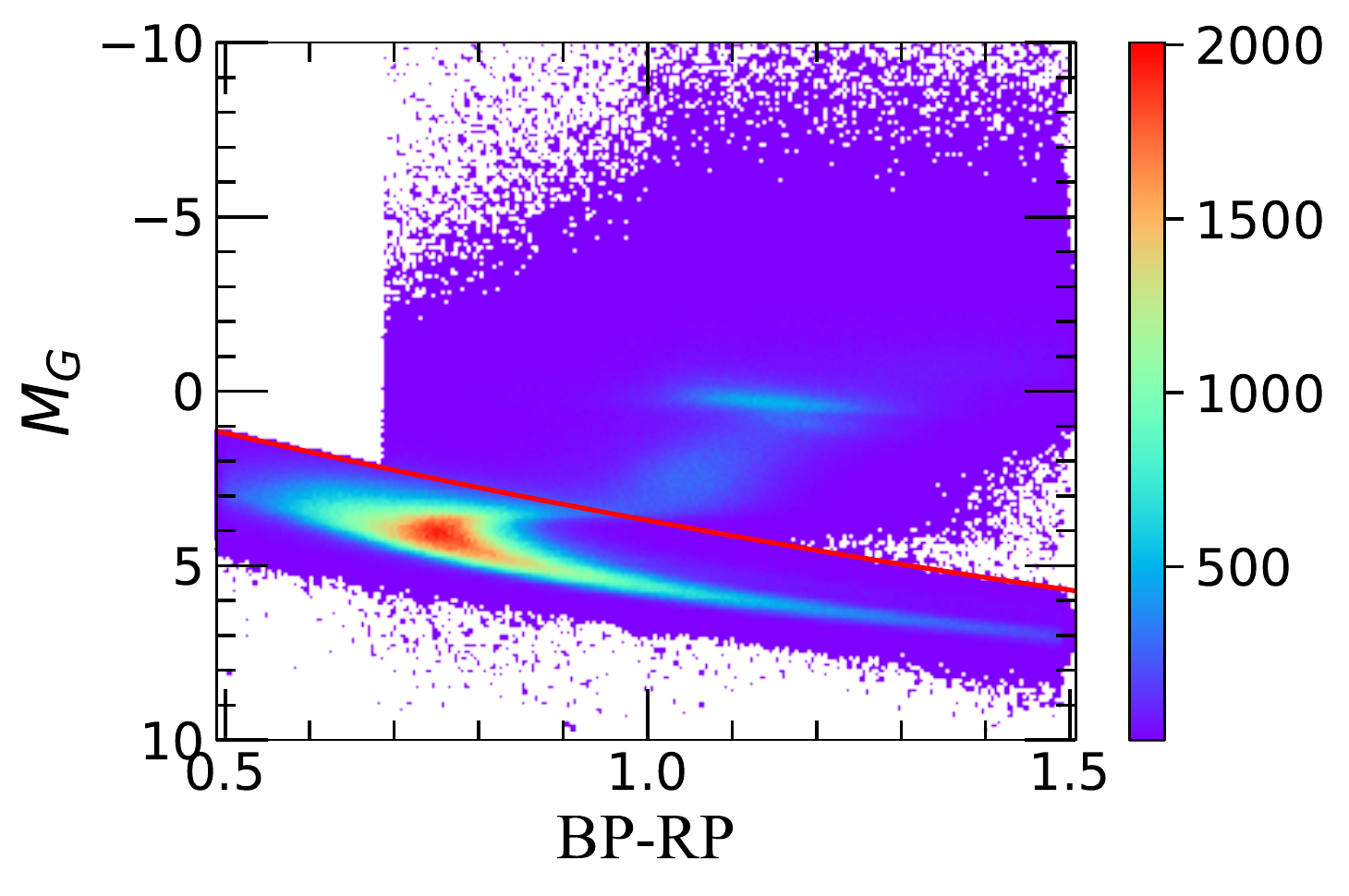}
    \caption{Color-magnitude  diagram  of  the  final sample. The colors indicate number densities, as shown in the color bar at right.  Dwarfs and giants are separated by the red line: $M_G = -{(BP-RP)}^2 + 6.5\times (BP-RP) - 1.8$.}
    \label{fig:final_HRD}
\end{figure}

In this section, we apply our methods to the entire {\it Gaia} data set in the applicable range of magnitude and and colors. 
We first select all stars of $10 < G \leq 16$ and  $|b| > 10^\circ$. Then, after the 
same cuts on phot\textunderscore bp\textunderscore rp\textunderscore excess\textunderscore factor and $BP-RP$ color as for the test sample are made, along with a cut on reddening $E(B-V) \leq 0.5$, 6,600,648 giants and 20,404,437 dwarfs remain in the final sample.  Their color-magnitude diagram is plotted in Figure \ref{fig:final_HRD}. 

Metallicities for stars in  the final sample are derived with our method (taking into account the corrections for bright stars given in Table 1), and the errors in metallicities are also estimated based on the photometric errors. The median errors are 0.29 dex and 0.34 dex for giants and dwarfs, respectively.
At $G = 14/15/16$, the typical errors are  0.22/0.35/0.61 dex for giants, and 0.21/0.34/0.58 dex for dwarfs, respectively. 

\begin{figure*}[htbp]
\centering
    \includegraphics[width=16cm]{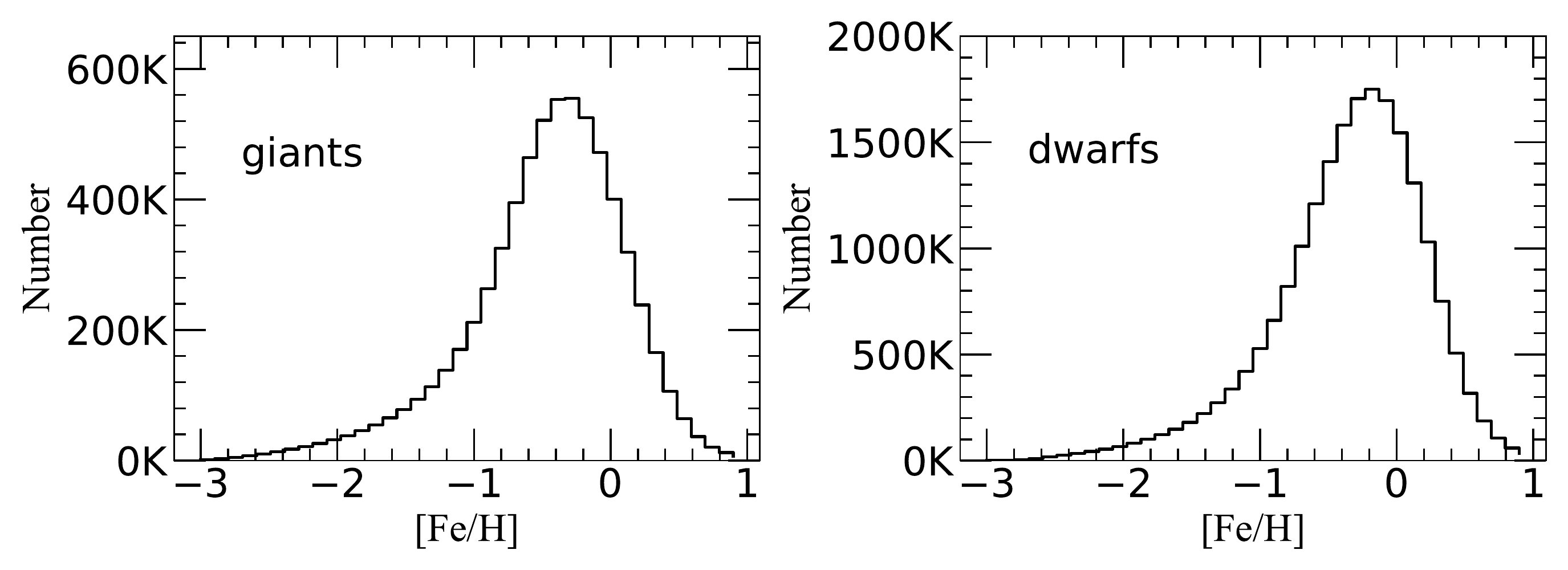}
    \caption{The [Fe/H] distributions of the final sample. The left and right panels are for the giants and dwarfs, respectively.}
    \label{fig:final_feh}
\end{figure*}

The [Fe/H] distributions are shown in Figure \ref{fig:final_feh}. 
For giants, there are 2.0, 15.7, 80.8, 19.2, and 2.0 per cent stars with [Fe/H] $\leq -2.0$, $\leq -1.0$, $\leq 0$, $> 0$, and $> +0.5$, respectively. 
For dwarfs, there are 1.1, 11.7, 73.3, 26.7, and 3.2 per cent stars with [Fe/H] $\leq -2.0$, $\leq -1.0$, $\leq 0$, $> 0$, and $> +0.5$, respectively. 

\begin{figure*}[htbp]
\centering
    \includegraphics[width=16cm]{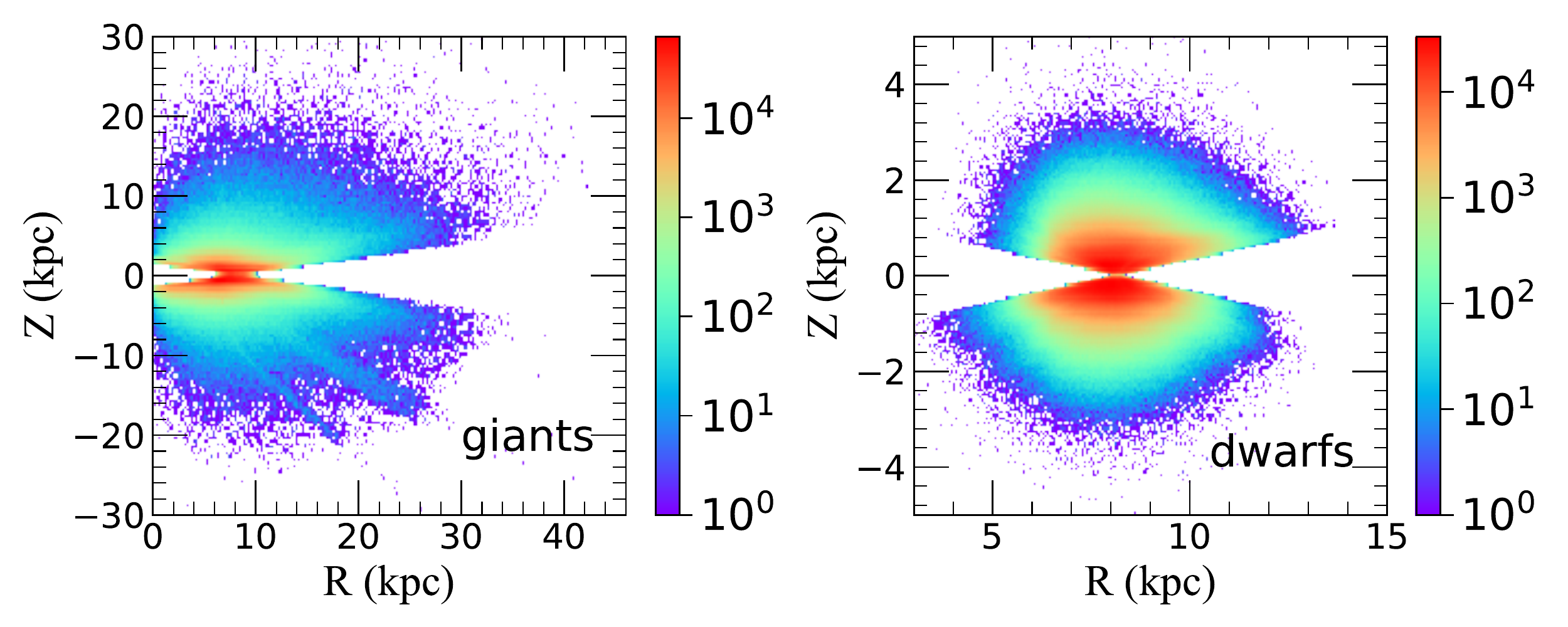}
    \caption{The spatial distributions in the Z--R plane of the final sample. The left and right panels are for the giants and dwarfs, respectively. The Sun is located at (Z, R) = (0, 8.178) kpc. The colors indicate number densities, as shown in the color bars. Note that the two 
    branches of stars visible in the lower right portion of the left panel are from the LMC and SMC. }
    \label{fig:final_R_Z}
\end{figure*}

\begin{figure*}[htbp]
\centering
    \includegraphics[width=16cm]{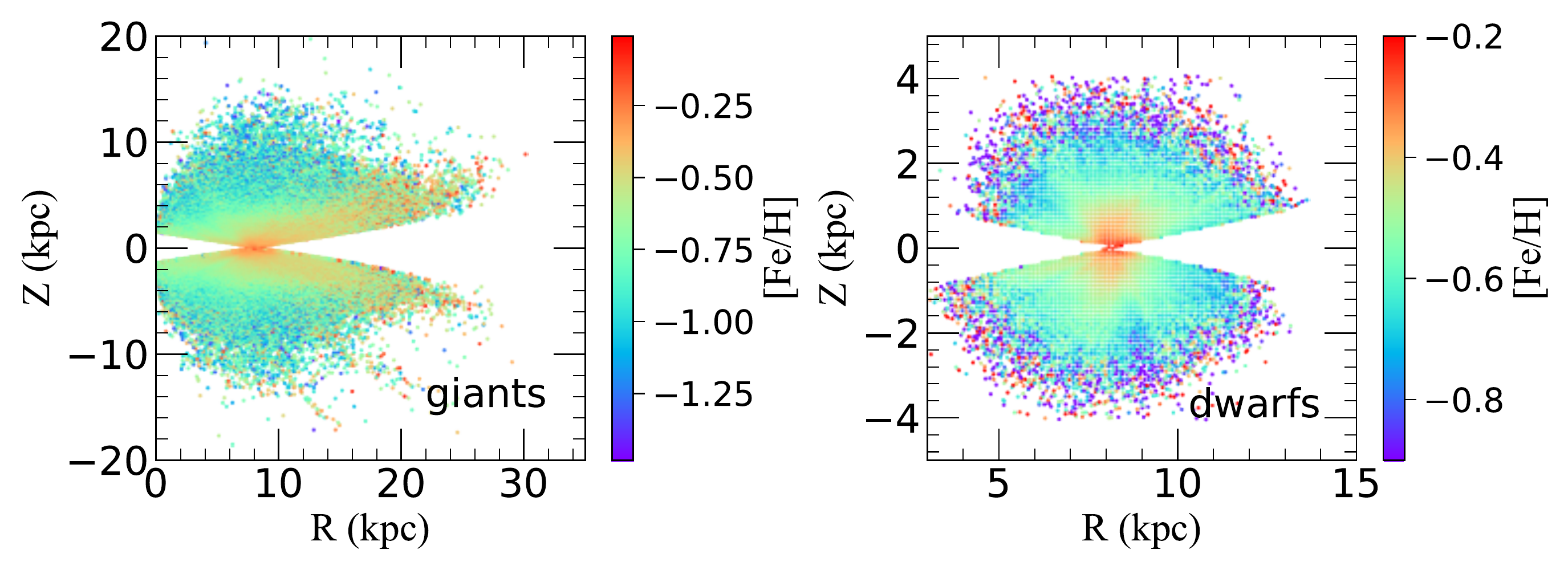}
    \caption{The spatial distributions in the Z--R plane of the final sample. The left and right panels are for the giants and dwarfs, respectively. The colors indicate [Fe/H], as shown in the color bars. The Sun is located at (Z, R) = (0, 8.178) kpc.}
    \label{fig:final_R_Z_feh}
\end{figure*}

The spatial distributions in the Z--R plane can be seen in Figure \ref{fig:final_R_Z}, where Z is the distance to the Galactic plane, and  R is the Galactocentric distance.
The giants and dwarfs are divided into different panels n the Z--R plane. 
Their median metallicities are plotted in Figure \ref{fig:final_R_Z_feh}. The vertical metallicity gradients of the disk stars
are clearly seen in both panels. 
The separation between disk stars and halo stars and the flaring of the Galactic disk are also 
clear in the left panel. 

Table \ref{table2} lists the columns contained in the final sample catalog.

\setlength{\tabcolsep}{4.7mm}{
\begin{table*}[htbp]
\footnotesize
\centering
\caption{Description of the Final Sample}
\begin{tabular}{lll}
\hline
\hline
Field & Description & Unit\\
\hline
source\_id & Unique source identifier for EDR3 (unique with a particular Data Release) & --\\
ra & Right ascension & deg\\
dec & Declination & deg\\
parallax & Parallax & mas\\
parallax\_error & Standard error of parallax & mas\\
pmra & Proper motion in right ascension direction & mas/year\\
pmra\_error & Standard error of proper motion in right ascension direction & mas/year\\
pmdec & Proper motion in declination direction & mas/year\\
pmdec\_error & Standard error of proper motion in declination direction & mas/year\\
ruwe & Renormalised unit weight error & --\\
phot\_g\_mean\_flux\_over\_error &  G-band mean flux divided by its error & --\\
phot\_g\_mean\_mag & G-band mean magnitude & --\\
phot\_bp\_mean\_flux\_over\_error & BP-band mean flux divided by its error & --\\
phot\_bp\_mean\_mag & Integrated BP-band mean magnitude & --\\
phot\_rp\_mean\_flux\_over\_error & RP-band mean flux divided by its error & --\\
phot\_rp\_mean\_mag & Integrated RP-band mean magnitude & --\\
phot\_bp\_rp\_excess\_factor & BP/RP excess factor & --\\
l & Galactic longitude & deg\\
b & Galactic latitude & deg\\
ebv & Value of E (B $-$ V ) from from the extinction map of SFD98 & --\\
correct\_bp\_rp & Intrinsic $BP - RP$ color after color correction of \cite{Niu2021EDR3} & --\\
correct\_bp\_g & Intrinsic $BP - G$ color after color correction of \cite{Niu2021EDR3} & --\\
type & Flag to indicate classifications of stars;  0 for dwarfs and 1 for giants & --\\
correct\_g\_rp & Intrinsic $G - RP$ color after color correction of \cite{Niu2021EDR3} & --\\
FeH\_Gaia & Photometric metallicity & -- \\
FeH\_Gaia\_error & Formal error of FeH\_Gaia  & dex\\
parallax\_corrected & Parallax corrected by \citet{Lindegren2021}& mas\\
\hline
\end{tabular}
\label{table2}
\end{table*}}

\section{Summary and future perspectives}\label{sec:con}

In this work, we combine LAMOST DR7 spectroscopic data and {\it Gaia} EDR3 photometric data to assemble 
high-quality giant (0.7 $< (BP-RP) <$ 1.4) and dwarf (0.5 $< (BP-RP) < $ 1.5) samples in the high Galactic latitude region, with precise corrections for magnitude-dependent systematic errors in the {\it Gaia} photometry and careful reddening corrections using empirical color- and reddening-dependent coefficients. 
The two samples are used to construct metallicity-dependent stellar loci of {\it Gaia} colors 
for giants and dwarfs, respectively. For a given $(BP-RP)$ color, a one dex change in [Fe/H] causes about a 5 mmag change in $(BP-G)$ color for solar-type stars. 
The metallicity-dependent loci are then used to determine stellar photometric metallicities from {\it Gaia} EDR3 colors.  Owing to the exquisite data quality of {\it Gaia},  we have achieved a typical precision of about 
0.2 dex for both giants and dwarfs, 
despite the very weak sensitivity of {\it Gaia} EDR3 colors on metallicity.
Our tests show that the method is valid for FGK stars with $G \leq 16$, [Fe/H] $\geq -2.5$, and $E(B-V) \leq 0.5$.  
Stars with fainter $G$ magnitudes, redder colors, lower metallicities, or larger reddening suffer from higher metallicity uncertainties. This work demonstrates the power of precision photometry in the determination of basic stellar parameters. 

With the enormous data volume of {\it Gaia}, our method has measured  metallicity estimates for about 27 million bright stars with $10 < G \leq 16$ across almost the entire sky, including over 6 million giant stars and 20 million dwarfs. This is the largest catalog of metallicity estimates for stars to date, and should prove useful for a number of studies. For example, one can construct a magnitude-limited sample of red giants to study the structure and chemistry of the inner Galactic halo and the outer disk, or a magnitude-limited sample of FGK dwarfs with known metallicities to better investigate the star-formation history of the Solar Neighborhood using observed {\it Gaia} color-magnitude diagrams (e.g., \citealt{Ruiz-Lara2020}).  The sample can also be used for the identification of candidate stars for subsequent high-resolution spectroscopic follow-up, or to select a pure sample of wide binaries (e.g., \citealt{tian2020,el2021}) 
based on stars with common proper motions and similar metallicities, or obtain metallicities for a very large fraction of the TESS targets and investigate the occurrence rate of hot Jupiters for metal-poor stars (e.g., \citealt{Boley2021}). The catalog is publicly available via \url{https://doi.org/10.12149/101081} 

Currently, our method uses the SFD reddening values to obtain metallicities from the {\it Gaia} colors.
For nearby stars in the low Galactic latitude region, where the SFD map is not valid, three-dimensional reddening maps (e.g., \citealt{Green2019, chen2019}) could be used instead, along with {\it Gaia} parallaxes. 
Alternatively, it is possible to derive reddening values and 
metallicities simultaneously, with help from infrared photometry obtained by 2MASS and WISE. Such explorations will be carried out in subsequent work.  Obviously, applications of our approach to future releases from {\it Gaia}, will enable improved metallicity estimates and allow us to probe more distant stars.

\begin{acknowledgments}
We acknowledge useful discussions with Dr. Xiaoting Fu.
This work is supported by the National Natural Science Foundation of China through the projects NSFC 12173007, 11603002, 11933004, National Key Basic R \& D Program of China via 2019YFA0405500, Beijing Normal University grant No. 310232102. T.C.B. acknowledges partial support from grant PHY 14-30152,
Physics Frontier Center/JINA Center for the Evolution
of the Elements (JINA-CEE), awarded by the US National
Science Foundation. His participation in this work was initiated
by conversations that took place during a visit to China
in 2019, supported by a PIFI Distinguished Scientist award
from the Chinese Academy of Science. We acknowledge the science research grants from the China Manned Space Project with NO. CMS-CSST-2021-A08 and CMS-CSST-2021-A09. 

This work has made use of data from the European Space Agency (ESA) mission {\it Gaia} (\url{https://www.cosmos.esa.int/gaia}), processed by the Gaia Data Processing and Analysis Consortium (DPAC, \url{https:// www.cosmos.esa.int/web/gaia/dpac/ consortium}).Funding for the DPAC has been provided by national institutions, in particular the institutions participating in the Gaia Multilateral Agreement. Guoshoujing Telescope (the Large Sky Area Multi-Object Fiber Spectroscopic Telescope LAMOST) is a National Major Scientific Project built by the Chinese Academy of Sciences. Funding for the project has been provided by the National Development and Reform Commission. LAMOST is operated and managed by the National Astronomical Observatories, Chinese Academy of Sciences.

\end{acknowledgments}

\bibliographystyle{aasjournal}
\bibliography{Gaia_metallicities}

\begin{thebibliography}{}
\expandafter\ifx\csname natexlab\endcsname\relax\def\natexlab#1{#1}\fi
\providecommand{\url}[1]{\href{#1}{#1}}
\providecommand{\dodoi}[1]{doi:~\href{http://doi.org/#1}{\nolinkurl{#1}}}
\providecommand{\doeprint}[1]{\href{http://ascl.net/#1}{\nolinkurl{http://ascl.net/#1}}}
\providecommand{\doarXiv}[1]{\href{https://arxiv.org/abs/#1}{\nolinkurl{https://arxiv.org/abs/#1}}}

\bibitem[{{Ahumada} {et~al.}(2020){Ahumada}, {Prieto}, {Almeida}, {Anders},
  {Anderson}, {Andrews}, {Anguiano}, {Arcodia}, {Armengaud}, {Aubert}, {Avila},
  {Avila-Reese}, {Badenes}, {Balland}, {Barger}, {Barrera-Ballesteros}, {Basu},
  {Bautista}, {Beaton}, {Beers}, {Benavides}, {Bender}, {Bernardi}, {Bershady},
  {Beutler}, {Bidin}, {Bird}, {Bizyaev}, {Blanc}, {Blanton}, {Boquien},
  {Borissova}, {Bovy}, {Brandt}, {Brinkmann}, {Brownstein}, {Bundy}, {Bureau},
  {Burgasser}, {Burtin}, {Cano-D{\'\i}az}, {Capasso}, {Cappellari}, {Carrera},
  {Chabanier}, {Chaplin}, {Chapman}, {Cherinka}, {Chiappini}, {Doohyun Choi},
  {Chojnowski}, {Chung}, {Clerc}, {Coffey}, {Comerford}, {Comparat}, {da
  Costa}, {Cousinou}, {Covey}, {Crane}, {Cunha}, {Ilha}, {Dai}, {Damsted},
  {Darling}, {Davidson}, {Davies}, {Dawson}, {De}, {de la Macorra}, {De Lee},
  {Queiroz}, {Deconto Machado}, {de la Torre}, {Dell'Agli}, {du Mas des
  Bourboux}, {Diamond-Stanic}, {Dillon}, {Donor}, {Drory}, {Duckworth},
  {Dwelly}, {Ebelke}, {Eftekharzadeh}, {Davis Eigenbrot}, {Elsworth},
  {Eracleous}, {Erfanianfar}, {Escoffier}, {Fan}, {Farr},
  {Fern{\'a}ndez-Trincado}, {Feuillet}, {Finoguenov}, {Fofie},
  {Fraser-McKelvie}, {Frinchaboy}, {Fromenteau}, {Fu}, {Galbany}, {Garcia},
  {Garc{\'\i}a-Hern{\'a}ndez}, {Oehmichen}, {Ge}, {Maia}, {Geisler}, {Gelfand},
  {Goddy}, {Gonzalez-Perez}, {Grabowski}, {Green}, {Grier}, {Guo}, {Guy},
  {Harding}, {Hasselquist}, {Hawken}, {Hayes}, {Hearty}, {Hekker}, {Hogg},
  {Holtzman}, {Horta}, {Hou}, {Hsieh}, {Huber}, {Hunt}, {Chitham}, {Imig},
  {Jaber}, {Angel}, {Johnson}, {Jones}, {J{\"o}nsson}, {Jullo}, {Kim},
  {Kinemuchi}, {Kirkpatrick}, {Kite}, {Klaene}, {Kneib}, {Kollmeier}, {Kong},
  {Kounkel}, {Krishnarao}, {Lacerna}, {Lan}, {Lane}, {Law}, {Le Goff}, {Leung},
  {Lewis}, {Li}, {Lian}, {Lin}, {Long}, {Longa-Pe{\~n}a}, {Lundgren}, {Lyke},
  {Ted Mackereth}, {MacLeod}, {Majewski}, {Manchado}, {Maraston}, {Martini},
  {Masseron}, {Masters}, {Mathur}, {McDermid}, {Merloni}, {Merrifield},
  {M{\'e}sz{\'a}ros}, {Miglio}, {Minniti}, {Minsley}, {Miyaji}, {Mohammad},
  {Mosser}, {Mueller}, {Muna}, {Mu{\~n}oz-Guti{\'e}rrez}, {Myers}, {Nadathur},
  {Nair}, {Nandra}, {do Nascimento}, {Nevin}, {Newman}, {Nidever}, {Nitschelm},
  {Noterdaeme}, {O'Connell}, {Olmstead}, {Oravetz}, {Oravetz}, {Osorio},
  {Pace}, {Padilla}, {Palanque-Delabrouille}, {Palicio}, {Pan}, {Pan},
  {Parker}, {Paviot}, {Peirani}, {Ram{\'r}ez}, {Penny}, {Percival},
  {Perez-Fournon}, {P{\'e}rez-R{\`a}fols}, {Petitjean}, {Pieri},
  {Pinsonneault}, {Poovelil}, {Povick}, {Prakash}, {Price-Whelan}, {Raddick},
  {Raichoor}, {Ray}, {Rembold}, {Rezaie}, {Riffel}, {Riffel}, {Rix}, {Robin},
  {Roman-Lopes}, {Rom{\'a}n-Z{\'u}{\~n}iga}, {Rose}, {Ross}, {Rossi},
  {Rowlands}, {Rubin}, {Salvato}, {S{\'a}nchez}, {S{\'a}nchez-Menguiano},
  {S{\'a}nchez-Gallego}, {Sayres}, {Schaefer}, {Schiavon}, {Schimoia},
  {Schlafly}, {Schlegel}, {Schneider}, {Schultheis}, {Schwope}, {Seo},
  {Serenelli}, {Shafieloo}, {Shamsi}, {Shao}, {Shen}, {Shetrone}, {Shirley},
  {Aguirre}, {Simon}, {Skrutskie}, {Slosar}, {Smethurst}, {Sobeck}, {Sodi},
  {Souto}, {Stark}, {Stassun}, {Steinmetz}, {Stello}, {Stermer},
  {Storchi-Bergmann}, {Streblyanska}, {Stringfellow}, {Stutz}, {Su{\'a}rez},
  {Sun}, {Taghizadeh-Popp}, {Talbot}, {Tayar}, {Thakar}, {Theriault}, {Thomas},
  {Thomas}, {Tinker}, {Tojeiro}, {Toledo}, {Tremonti}, {Troup}, {Tuttle},
  {Unda-Sanzana}, {Valentini}, {Vargas-Gonz{\'a}lez}, {Vargas-Maga{\~n}a},
  {V{\'a}zquez-Mata}, {Vivek}, {Wake}, {Wang}, {Weaver}, {Weijmans}, {Wild},
  {Wilson}, {Wilson}, {Wolthuis}, {Wood-Vasey}, {Yan}, {Yang}, {Y{\`e}che},
  {Zamora}, {Zarrouk}, {Zasowski}, {Zhang}, {Zhao}, {Zhao}, {Zheng}, {Zheng},
  {Zhu}, \& {Zou}}]{SDSSfeh}
{Ahumada}, R., {Prieto}, C.~A., {Almeida}, A., {et~al.} 2020, \apjs, 249, 3,
  \dodoi{10.3847/1538-4365/ab929e}

\bibitem[{{Allende Prieto} {et~al.}(2008){Allende Prieto}, {Sivarani}, {Beers},
  {Lee}, {Koesterke}, {Shetrone}, {Sneden}, {Lambert}, {Wilhelm}, {Rockosi},
  {Lai}, {Yanny}, {Ivans}, {Johnson}, {Aoki}, {Bailer-Jones}, \& {Re
  Fiorentin}}]{Lee2008c}
{Allende Prieto}, C., {Sivarani}, T., {Beers}, T.~C., {et~al.} 2008, \aj, 136,
  2070, \dodoi{10.1088/0004-6256/136/5/2070}

\bibitem[{{An} \& {Beers}(2020)}]{an2020}
{An}, D., \& {Beers}, T.~C. 2020, \apj, 897, 39,
  \dodoi{10.3847/1538-4357/ab8d39}

\bibitem[{{An} \& {Beers}(2021{\natexlab{a}})}]{an2021a}
---. 2021{\natexlab{a}}, \apj, 907, 101, \dodoi{10.3847/1538-4357/abccd2}

\bibitem[{{An} \& {Beers}(2021{\natexlab{b}})}]{an2021b}
---. 2021{\natexlab{b}}, \apj, 918, 74, \dodoi{10.3847/1538-4357/ac07a4}

\bibitem[{{Arce} \& {Goodman}(1999)}]{Arce1999}
{Arce}, H.~G., \& {Goodman}, A.~A. 1999, \apjl, 512, L135,
  \dodoi{10.1086/311885}

\bibitem[{{Beers} \& {Christlieb}(2005)}]{beers2005}
{Beers}, T.~C., \& {Christlieb}, N. 2005, Highlights of Astronomy, 13, 579

\bibitem[{{Beers} {et~al.}(2014){Beers}, {Norris}, {Placco}, {Lee}, {Rossi},
  {Carollo}, \& {Masseron}}]{Beers2014}
{Beers}, T.~C., {Norris}, J.~E., {Placco}, V.~M., {et~al.} 2014, \apj, 794, 58,
  \dodoi{10.1088/0004-637X/794/1/58}

\bibitem[{{Beers} {et~al.}(2017){Beers}, {Placco}, {Carollo}, {Rossi}, {Lee},
  {Frebel}, {Norris}, {Dietz}, \& {Masseron}}]{Beers2017}
{Beers}, T.~C., {Placco}, V.~M., {Carollo}, D., {et~al.} 2017, \apj, 835, 81,
  \dodoi{10.3847/1538-4357/835/1/81}

\bibitem[{{Boley} {et~al.}(2021){Boley}, {Wang}, {Zinn}, {Collins}, {Collins},
  {Gan}, \& {Li}}]{Boley2021}
{Boley}, K.~M., {Wang}, J., {Zinn}, J.~C., {et~al.} 2021, \aj, 162, 85,
  \dodoi{10.3847/1538-3881/ac0e2d}

\bibitem[{{Bonifacio} {et~al.}(2021){Bonifacio}, {Monaco}, {Salvadori},
  {Caffau}, {Spite}, {Sbordone}, {Spite}, {Ludwig}, {Di Matteo}, {Haywood},
  {Fran{\c{c}}ois}, {Koch-Hansen}, {Christlieb}, \& {Zaggia}}]{bonifacio2021}
{Bonifacio}, P., {Monaco}, L., {Salvadori}, S., {et~al.} 2021, \aap, 651, A79,
  \dodoi{10.1051/0004-6361/202140816}

\bibitem[{Brown {et~al.}(2021)Brown, Vallenari, Prusti, de~Bruijne, Babusiaux,
  Biermann, Creevey, Evans, Eyer, \& et~al.}]{GaiaEDR32021}
Brown, A. G.~A., Vallenari, A., Prusti, T., {et~al.} 2021, Astronomy \&
  Astrophysics, 650, C3, \dodoi{10.1051/0004-6361/202039657e}

\bibitem[{Buder {et~al.}(2018)Buder, Asplund, Duong, Kos, Lind, Ness, Sharma,
  Bland-Hawthorn, Casey, De Silva, D’Orazi, Freeman, Lewis, Lin, Martell,
  Schlesinger, Simpson, Zucker, Zwitter, Amarsi, Anguiano, Carollo, Casagrande,
  Čotar, Cottrell, Da Costa, Gao, Hayden, Horner, Ireland, Kafle, Munari,
  Nataf, Nordlander, Stello, Ting~(丁源森), Traven, Watson, Wittenmyer,
  Wyse, Yong, Zinn, Žerjal, \& collaboration}]{GALAH2018}
Buder, S., Asplund, M., Duong, L., {et~al.} 2018, Monthly Notices of the Royal
  Astronomical Society, 478, 4513, \dodoi{10.1093/mnras/sty1281}

\bibitem[{{Buder} {et~al.}(2021){Buder}, {Sharma}, {Kos}, {Amarsi},
  {Nordlander}, {Lind}, {Martell}, {Asplund}, {Bland-Hawthorn}, {Casey}, {de
  Silva}, {D'Orazi}, {Freeman}, {Hayden}, {Lewis}, {Lin}, {Schlesinger},
  {Simpson}, {Stello}, {Zucker}, {Zwitter}, {Beeson}, {Buck}, {Casagrande},
  {Clark}, {{\v{C}}otar}, {da Costa}, {de Grijs}, {Feuillet}, {Horner},
  {Kafle}, {Khanna}, {Kobayashi}, {Liu}, {Montet}, {Nandakumar}, {Nataf},
  {Ness}, {Spina}, {Tepper-Garc{\'\i}a}, {Ting}, {Traven},
  {Vogrin{\v{c}}i{\v{c}}}, {Wittenmyer}, {Wyse}, {{\v{Z}}erjal}, \& {Galah
  Collaboration}}]{GALAHfeh}
{Buder}, S., {Sharma}, S., {Kos}, J., {et~al.} 2021, \mnras,
  \dodoi{10.1093/mnras/stab1242}

\bibitem[{{Cantat-Gaudin} {et~al.}(2018){Cantat-Gaudin}, {Jordi}, {Vallenari},
  {Bragaglia}, {Balaguer-N{\'u}{\~n}ez}, {Soubiran}, {Bossini}, {Moitinho},
  {Castro-Ginard}, {Krone-Martins}, {Casamiquela}, {Sordo}, \&
  {Carrera}}]{cluster}
{Cantat-Gaudin}, T., {Jordi}, C., {Vallenari}, A., {et~al.} 2018, \aap, 618,
  A93, \dodoi{10.1051/0004-6361/201833476}

\bibitem[{{Casagrande} {et~al.}(2011){Casagrande}, {Sch{\"o}nrich}, {Asplund},
  {Cassisi}, {Ram{\'\i}rez}, {Mel{\'e}ndez}, {Bensby}, \&
  {Feltzing}}]{Casagrand2011}
{Casagrande}, L., {Sch{\"o}nrich}, R., {Asplund}, M., {et~al.} 2011, \aap, 530,
  A138, \dodoi{10.1051/0004-6361/201016276}

\bibitem[{{Casagrande} {et~al.}(2019){Casagrande}, {Wolf}, {Mackey},
  {Nordlander}, {Yong}, \& {Bessell}}]{Casagrande2019}
{Casagrande}, L., {Wolf}, C., {Mackey}, A.~D., {et~al.} 2019, \mnras, 482,
  2770, \dodoi{10.1093/mnras/sty2878}

\bibitem[{{Casagrande} {et~al.}(2021){Casagrande}, {Lin}, {Rains}, {Liu},
  {Buder}, {Horner}, {Asplund}, {Lewis}, {Martell}, {Nordlander}, {Stello},
  {Ting}, {Wittenmyer}, {Bland-Hawthorn}, {Casey}, {De Silva}, {D'Orazi},
  {Freeman}, {Hayden}, {Kos}, {Lind}, {Schlesinger}, {Sharma}, {Simpson},
  {Zucker}, \& {Zwitter}}]{Casagrande2021}
{Casagrande}, L., {Lin}, J., {Rains}, A.~D., {et~al.} 2021, \mnras, 507, 2684,
  \dodoi{10.1093/mnras/stab2304}

\bibitem[{{Chen} {et~al.}(2019){Chen}, {Huang}, {Yuan}, {Wang}, {Fan}, {Xiang},
  {Zhang}, {Tian}, \& {Liu}}]{chen2019}
{Chen}, B.~Q., {Huang}, Y., {Yuan}, H.~B., {et~al.} 2019, \mnras, 483, 4277,
  \dodoi{10.1093/mnras/sty3341}

\bibitem[{{El-Badry} {et~al.}(2021){El-Badry}, {Rix}, \& {Heintz}}]{el2021}
{El-Badry}, K., {Rix}, H.-W., \& {Heintz}, T.~M. 2021, \mnras, 506, 2269,
  \dodoi{10.1093/mnras/stab323}

\bibitem[{{Gaia Collaboration} {et~al.}(2021){Gaia Collaboration}, {Antoja},
  {McMillan}, {Kordopatis}, {Ramos}, {Helmi}, {Balbinot}, {Cantat-Gaudin},
  {Chemin}, {Figueras}, {Jordi}, {Khanna}, {Romero-G{\'o}mez}, {Seabroke},
  {Brown}, {Vallenari}, {Prusti}, {de Bruijne}, {Babusiaux}, {Biermann},
  {Creevey}, {Evans}, {Eyer}, {Hutton}, {Jansen}, {Klioner}, {Lammers},
  {Lindegren}, {Luri}, {Mignard}, {Panem}, {Pourbaix}, {Randich}, {Sartoretti},
  {Soubiran}, {Walton}, {Arenou}, {Bailer-Jones}, {Bastian}, {Cropper},
  {Drimmel}, {Katz}, {Lattanzi}, {van Leeuwen}, {Bakker}, {Casta{\~n}eda}, {De
  Angeli}, {Ducourant}, {Fabricius}, {Fouesneau}, {Fr{\'e}mat}, {Guerra},
  {Guerrier}, {Guiraud}, {Jean-Antoine Piccolo}, {Masana}, {Messineo},
  {Mowlavi}, {Nicolas}, {Nienartowicz}, {Pailler}, {Panuzzo}, {Riclet}, {Roux},
  {Sordo}, {Tanga}, {Th{\'e}venin}, {Gracia-Abril}, {Portell}, {Teyssier},
  {Altmann}, {Andrae}, {Bellas-Velidis}, {Benson}, {Berthier}, {Blomme},
  {Brugaletta}, {Burgess}, {Busso}, {Carry}, {Cellino}, {Cheek}, {Clementini},
  {Damerdji}, {Davidson}, {Delchambre}, {Dell'Oro},
  {Fern{\'a}ndez-Hern{\'a}ndez}, {Galluccio}, {Garc{\'\i}a-Lario},
  {Garcia-Reinaldos}, {Gonz{\'a}lez-N{\'u}{\~n}ez}, {Gosset}, {Haigron},
  {Halbwachs}, {Hambly}, {Harrison}, {Hatzidimitriou}, {Heiter},
  {Hern{\'a}ndez}, {Hestroffer}, {Hodgkin}, {Holl}, {Jan{\ss}en}, {Jevardat de
  Fombelle}, {Jordan}, {Krone-Martins}, {Lanzafame}, {L{\"o}ffler}, {Lorca},
  {Manteiga}, {Marchal}, {Marrese}, {Moitinho}, {Mora}, {Muinonen}, {Osborne},
  {Pancino}, {Pauwels}, {Recio-Blanco}, {Richards}, {Riello}, {Rimoldini},
  {Robin}, {Roegiers}, {Rybizki}, {Sarro}, {Siopis}, {Smith}, {Sozzetti},
  {Ulla}, {Utrilla}, {van Leeuwen}, {van Reeven}, {Abbas}, {Abreu Aramburu},
  {Accart}, {Aerts}, {Aguado}, {Ajaj}, {Altavilla}, {{\'A}lvarez}, {{\'A}lvarez
  Cid-Fuentes}, {Alves}, {Anderson}, {Varela}, {Audard}, {Baines}, {Baker},
  {Balaguer-N{\'u}{\~n}ez}, {Balog}, {Barache}, {Barbato}, {Barros}, {Barstow},
  {Bartolom{\'e}}, {Bassilana}, {Bauchet}, {Baudesson-Stella}, {Becciani},
  {Bellazzini}, {Bernet}, {Bertone}, {Bianchi}, {Blanco-Cuaresma}, {Boch},
  {Bombrun}, {Bossini}, {Bouquillon}, {Bragaglia}, {Bramante}, {Breedt},
  {Bressan}, {Brouillet}, {Bucciarelli}, {Burlacu}, {Busonero}, {Butkevich},
  {Buzzi}, {Caffau}, {Cancelliere}, {C{\'a}novas}, {Carballo}, {Carlucci},
  {Carnerero}, {Carrasco}, {Casamiquela}, {Castellani}, {Castro-Ginard},
  {Castro Sampol}, {Chaoul}, {Charlot}, {Chiavassa}, {Cioni}, {Comoretto},
  {Cooper}, {Cornez}, {Cowell}, {Crifo}, {Crosta}, {Crowley}, {Dafonte},
  {Dapergolas}, {David}, {David}, {de Laverny}, {De Luise}, {De March}, {De
  Ridder}, {de Souza}, {de Teodoro}, {de Torres}, {del Peloso}, {del Pozo},
  {Delgado}, {Delgado}, {Delisle}, {Di Matteo}, {Diakite}, {Diener},
  {Distefano}, {Dolding}, {Eappachen}, {Enke}, {Esquej}, {Fabre}, {Fabrizio},
  {Faigler}, {Fedorets}, {Fernique}, {Fienga}, {Fouron}, {Fragkoudi}, {Fraile},
  {Franke}, {Gai}, {Garabato}, {Garcia-Gutierrez}, {Garc{\'\i}a-Torres},
  {Garofalo}, {Gavras}, {Gerlach}, {Geyer}, {Giacobbe}, {Gilmore}, {Girona},
  {Giuffrida}, {Gomez}, {Gonzalez-Santamaria}, {Gonz{\'a}lez-Vidal}, {Granvik},
  {Guti{\'e}rrez-S{\'a}nchez}, {Guy}, {Hauser}, {Haywood}, {Hidalgo}, {Hilger},
  {H{\l}adczuk}, {Hobbs}, {Holland}, {Huckle}, {Jasniewicz}, {Jonker},
  {Juaristi Campillo}, {Julbe}, {Karbevska}, {Kervella}, {Kochoska},
  {Kontizas}, {Korn}, {Kostrzewa-Rutkowska}, {Kruszy{\'n}ska}, {Lambert},
  {Lanza}, {Lasne}, {Le Campion}, {Le Fustec}, {Lebreton}, {Lebzelter},
  {Leccia}, {Leclerc}, {Lecoeur-Taibi}, {Liao}, {Licata}, {Lindstr{\o}m},
  {Lister}, {Livanou}, {Lobel}, {Madrero Pardo}, {Managau}, {Mann}, {Marchant},
  {Marconi}, {Marcos Santos}, {Marinoni}, {Marocco}, {Marshall}, {Martin Polo},
  {Mart{\'\i}n-Fleitas}, {Masip}, {Massari}, {Mastrobuono-Battisti}, {Mazeh},
  {Messina}, {Michalik}, {Millar}, {Mints}, {Molina}, {Molinaro}, {Moln{\'a}r},
  {Montegriffo}, {Mor}, {Morbidelli}, {Morel}, {Morris}, {Mulone}, {Munoz},
  {Muraveva}, {Murphy}, {Musella}, {Noval}, {Ord{\'e}novic}, {Orr{\`u}},
  {Osinde}, {Pagani}, {Pagano}, {Palaversa}, {Palicio}, {Panahi}, {Pawlak},
  {Pe{\~n}alosa Esteller}, {Penttil{\"a}}, {Piersimoni}, {Pineau}, {Plachy},
  {Plum}, {Poggio}, {Poretti}, {Poujoulet}, {Pr{\v{s}}a}, {Pulone}, {Racero},
  {Ragaini}, {Rainer}, {Raiteri}, {Rambaux}, {Ramos-Lerate}, {Re Fiorentin},
  {Regibo}, {Reyl{\'e}}, {Ripepi}, {Riva}, {Rixon}, {Robichon}, {Robin},
  {Roelens}, {Rohrbasser}, {Rowell}, {Royer}, {Rybicki}, {Sadowski},
  {Sagrist{\`a} Sell{\'e}s}, {Sahlmann}, {Salgado}, {Salguero}, {Samaras},
  {Sanchez Gimenez}, {Sanna}, {Santove{\~n}a}, {Sarasso}, {Schultheis},
  {Sciacca}, {Segol}, {Segovia}, {S{\'e}gransan}, {Semeux}, {Siddiqui},
  {Siebert}, {Siltala}, {Slezak}, {Smart}, {Solano}, {Solitro}, {Souami},
  {Souchay}, {Spagna}, {Spoto}, {Steele}, {Steidelm{\"u}ller}, {Stephenson},
  {S{\"u}veges}, {Szabados}, {Szegedi-Elek}, {Taris}, {Tauran}, {Taylor},
  {Teixeira}, {Thuillot}, {Tonello}, {Torra}, {Torra}, {Turon}, {Unger},
  {Vaillant}, {van Dillen}, {Vanel}, {Vecchiato}, {Viala}, {Vicente},
  {Voutsinas}, {Weiler}, {Wevers}, {Wyrzykowski}, {Yoldas}, {Yvard}, {Zhao},
  {Zorec}, {Zucker}, {Zurbach}, \& {Zwitter}}]{Antoja2021}
{Gaia Collaboration}, {Antoja}, T., {McMillan}, P.~J., {et~al.} 2021, \aap,
  649, A8, \dodoi{10.1051/0004-6361/202039714}

\bibitem[{{Garc{\'\i}a P{\'e}rez} {et~al.}(2016){Garc{\'\i}a P{\'e}rez},
  {Allende Prieto}, {Holtzman}, {Shetrone}, {M{\'e}sz{\'a}ros}, {Bizyaev},
  {Carrera}, {Cunha}, {Garc{\'\i}a-Hern{\'a}ndez}, {Johnson}, {Majewski},
  {Nidever}, {Schiavon}, {Shane}, {Smith}, {Sobeck}, {Troup}, {Zamora},
  {Weinberg}, {Bovy}, {Eisenstein}, {Feuillet}, {Frinchaboy}, {Hayden},
  {Hearty}, {Nguyen}, {O'Connell}, {Pinsonneault}, {Wilson}, \&
  {Zasowski}}]{Garcia2016}
{Garc{\'\i}a P{\'e}rez}, A.~E., {Allende Prieto}, C., {Holtzman}, J.~A.,
  {et~al.} 2016, \aj, 151, 144, \dodoi{10.3847/0004-6256/151/6/144}

\bibitem[{{Green} {et~al.}(2019){Green}, {Schlafly}, {Zucker}, {Speagle}, \&
  {Finkbeiner}}]{Green2019}
{Green}, G.~M., {Schlafly}, E., {Zucker}, C., {Speagle}, J.~S., \&
  {Finkbeiner}, D. 2019, \apj, 887, 93, \dodoi{10.3847/1538-4357/ab5362}

\bibitem[{{Huang} {et~al.}(2021{\natexlab{a}}){Huang}, {Yuan}, {Beers}, \&
  {Zhang}}]{huangoffest}
{Huang}, Y., {Yuan}, H., {Beers}, T.~C., \& {Zhang}, H. 2021{\natexlab{a}},
  \apjl, 910, L5, \dodoi{10.3847/2041-8213/abe69a}

\bibitem[{{Huang} {et~al.}(2019){Huang}, {Chen}, {Yuan}, {Zhang}, {Xiang},
  {Wang}, {Wang}, {Wolf}, {Liu}, \& {Liu}}]{Huang2019}
{Huang}, Y., {Chen}, B.~Q., {Yuan}, H.~B., {et~al.} 2019, \apjs, 243, 7,
  \dodoi{10.3847/1538-4365/ab1f72}

\bibitem[{{Huang} {et~al.}(2021{\natexlab{b}}){Huang}, {Beers}, {Wolf}, {Lee},
  {Onken}, {Yuan}, {Shank}, {Zhang}, {Wang}, {Shi}, \& {Fan}}]{huang2021}
{Huang}, Y., {Beers}, T.~C., {Wolf}, C., {et~al.} 2021{\natexlab{b}}, arXiv
  e-prints, arXiv:2104.14154.
\newblock \doarXiv{2104.14154}

\bibitem[{{Ivezi{\'c}} {et~al.}(2008){Ivezi{\'c}}, {Sesar}, {Juri{\'c}},
  {Bond}, {Dalcanton}, {Rockosi}, {Yanny}, {Newberg}, {Beers}, {Allende
  Prieto}, {Wilhelm}, {Lee}, {Sivarani}, {Norris}, {Bailer-Jones}, {Re
  Fiorentin}, {Schlegel}, {Uomoto}, {Lupton}, {Knapp}, {Gunn}, {Covey}, {Allyn
  Smith}, {Miknaitis}, {Doi}, {Tanaka}, {Fukugita}, {Kent}, {Finkbeiner},
  {Munn}, {Pier}, {Quinn}, {Hawley}, {Anderson}, {Kiuchi}, {Chen}, {Bushong},
  {Sohi}, {Haggard}, {Kimball}, {Barentine}, {Brewington}, {Harvanek},
  {Kleinman}, {Krzesinski}, {Long}, {Nitta}, {Snedden}, {Lee}, {Harris},
  {Brinkmann}, {Schneider}, \& {York}}]{Ivezic2008}
{Ivezi{\'c}}, {\v{Z}}., {Sesar}, B., {Juri{\'c}}, M., {et~al.} 2008, \apj, 684,
  287, \dodoi{10.1086/589678}

\bibitem[{{J{\"o}nsson} {et~al.}(2020){J{\"o}nsson}, {Holtzman}, {Allende
  Prieto}, {Cunha}, {Garc{\'\i}a-Hern{\'a}ndez}, {Hasselquist}, {Masseron},
  {Osorio}, {Shetrone}, {Smith}, {Stringfellow}, {Bizyaev}, {Edvardsson},
  {Majewski}, {M{\'e}sz{\'a}ros}, {Souto}, {Zamora}, {Beaton}, {Bovy}, {Donor},
  {Pinsonneault}, {Poovelil}, \& {Sobeck}}]{RAVEfeh}
{J{\"o}nsson}, H., {Holtzman}, J.~A., {Allende Prieto}, C., {et~al.} 2020, \aj,
  160, 120, \dodoi{10.3847/1538-3881/aba592}

\bibitem[{{Lee} {et~al.}(2008{\natexlab{a}}){Lee}, {Beers}, {Sivarani},
  {Allende Prieto}, {Koesterke}, {Wilhelm}, {Re Fiorentin}, {Bailer-Jones},
  {Norris}, {Rockosi}, {Yanny}, {Newberg}, {Covey}, {Zhang}, \&
  {Luo}}]{Lee2008a}
{Lee}, Y.~S., {Beers}, T.~C., {Sivarani}, T., {et~al.} 2008{\natexlab{a}}, \aj,
  136, 2022, \dodoi{10.1088/0004-6256/136/5/2022}

\bibitem[{{Lee} {et~al.}(2008{\natexlab{b}}){Lee}, {Beers}, {Sivarani},
  {Johnson}, {An}, {Wilhelm}, {Allende Prieto}, {Koesterke}, {Re Fiorentin},
  {Bailer-Jones}, {Norris}, {Yanny}, {Rockosi}, {Newberg}, {Cudworth}, \&
  {Pan}}]{Lee2008b}
---. 2008{\natexlab{b}}, \aj, 136, 2050, \dodoi{10.1088/0004-6256/136/5/2050}

\bibitem[{{Lee} {et~al.}(2011){Lee}, {Beers}, {An}, {Ivezi{\'c}}, {Just},
  {Rockosi}, {Morrison}, {Johnson}, {Sch{\"o}nrich}, {Bird}, {Yanny},
  {Harding}, \& {Rocha-Pinto}}]{Lee2011}
{Lee}, Y.~S., {Beers}, T.~C., {An}, D., {et~al.} 2011, \apj, 738, 187,
  \dodoi{10.1088/0004-637X/738/2/187}

\bibitem[{{Lee} {et~al.}(2013){Lee}, {Beers}, {Masseron}, {Plez}, {Rockosi},
  {Sobeck}, {Yanny}, {Lucatello}, {Sivarani}, {Placco}, \& {Carollo}}]{Lee2013}
{Lee}, Y.~S., {Beers}, T.~C., {Masseron}, T., {et~al.} 2013, \aj, 146, 132,
  \dodoi{10.1088/0004-6256/146/5/132}

\bibitem[{{Lindegren} {et~al.}(2021){Lindegren}, {Bastian}, {Biermann},
  {Bombrun}, {de Torres}, {Gerlach}, {Geyer}, {Hern{\'a}ndez}, {Hilger},
  {Hobbs}, {Klioner}, {Lammers}, {McMillan}, {Ramos-Lerate},
  {Steidelm{\"u}ller}, {Stephenson}, \& {van Leeuwen}}]{Lindegren2021}
{Lindegren}, L., {Bastian}, U., {Biermann}, M., {et~al.} 2021, \aap, 649, A4,
  \dodoi{10.1051/0004-6361/202039653}

\bibitem[{Luo {et~al.}(2015)Luo, Zhao, Zhao, Deng, Liu, Jing, Wang, Zhang, Shi,
  Cui, Chu, Li, Bai, Wu, Cai, Cao, Cao, Carlin, Chen, Chen, Chen, Chen, Chen,
  Chen, Chen, Christlieb, Chu, Cui, Dong, Du, Fan, Feng, Fu, Gao, Gong, Gu,
  Guo, Han, He, Hou, Hou, Hou, Hu, Hu, Hu, Huo, Jia, Jiang, Jiang, Jiang, Jin,
  Kong, Kong, Lei, Li, Li, Li, Li, Li, Li, Li, Li, Li, Li, Li, Li, Liang, Lin,
  Liu, Liu, Liu, Liu, Lu, Luo, Mao, Newberg, Ni, Qi, Qi, Shen, Shi, Song, Song,
  Su, Su, Tang, Tao, Tian, Wang, Wang, Wang, Wang, Wang, Wang, Wang, Wang,
  Wang, Wang, Wang, Wang, Wang, Wang, Wang, Wang, Wang, Wang, Wang, Wang, Wei,
  Wei, Wu, Wu, Wu, Wu, Xing, Xu, Xu, Xu, Yan, Yang, Yang, Yang, Yang, Yao, Yu,
  Yuan, Yuan, Yuan, Yuan, Zhai, Zhang, Zhang, Zhang, Zhang, Zhang, Zhang,
  Zhang, Zhang, Zhao, Zhou, Zhou, Zhu, Zhu, Zou, \& Zuo}]{LAMOSTDR5}
Luo, A.-L., Zhao, Y.-H., Zhao, G., {et~al.} 2015, Research in Astronomy and
  Astrophysics, 15, 1095, \dodoi{10.1088/1674-4527/15/8/002}

\bibitem[{{Ma{\'\i}z Apell{\'a}niz} \& {Weiler}(2018)}]{ma2018}
{Ma{\'\i}z Apell{\'a}niz}, J., \& {Weiler}, M. 2018, \aap, 619, A180,
  \dodoi{10.1051/0004-6361/201834051}

\bibitem[{{Majewski} {et~al.}(2016){Majewski}, {APOGEE Team}, \& {APOGEE-2
  Team}}]{Majewski2016}
{Majewski}, S.~R., {APOGEE Team}, \& {APOGEE-2 Team}. 2016, Astronomische
  Nachrichten, 337, 863, \dodoi{10.1002/asna.201612387}

\bibitem[{{Niu} {et~al.}(2021{\natexlab{a}}){Niu}, {Yuan}, \&
  {Liu}}]{niu2021DR2}
{Niu}, Z., {Yuan}, H., \& {Liu}, J. 2021{\natexlab{a}}, \apj, 909, 48,
  \dodoi{10.3847/1538-4357/abdbac}

\bibitem[{{Niu} {et~al.}(2021{\natexlab{b}}){Niu}, {Yuan}, \&
  {Liu}}]{Niu2021EDR3}
---. 2021{\natexlab{b}}, \apjl, 908, L14, \dodoi{10.3847/2041-8213/abe1c2}

\bibitem[{Niu {et~al.}(2021)Niu, Yuan, Wang, \& Liu}]{niu2021binary}
Niu, Z., Yuan, H., Wang, S., \& Liu, J. 2021, Binary fractions of G and K dwarf
  stars based on the Gaia EDR3 and LAMOST DR5: impacts of the chemical
  abundances.
\newblock \doarXiv{2109.04031}

\bibitem[{{Onken} {et~al.}(2019){Onken}, {Wolf}, {Bessell}, {Chang}, {Da
  Costa}, {Luvaul}, {Mackey}, {Schmidt}, \& {Shao}}]{onken2019}
{Onken}, C.~A., {Wolf}, C., {Bessell}, M.~S., {et~al.} 2019, \pasa, 36, e033,
  \dodoi{10.1017/pasa.2019.27}

\bibitem[{Peng {et~al.}(2013)Peng, Du, Wu, Ma, \& Zhou}]{Peng2013}
Peng, X., Du, C., Wu, Z., Ma, J., \& Zhou, X. 2013, Monthly Notices of the
  Royal Astronomical Society, 434, 3165, \dodoi{10.1093/mnras/stt1232}

\bibitem[{{Ren} {et~al.}(2020){Ren}, {Raddi}, {Rebassa-Mansergas}, {Hernandez},
  {Parsons}, {Irawati}, {Rittipruk}, {Schreiber}, {G{\"a}nsicke}, {Torres},
  {Wang}, {Zhang}, {Zhao}, {Zhou}, {Han}, {Wang}, {Liu}, {Liu}, {Wang},
  {Zheng}, {Wang}, {Zhao}, {Cui}, {Shi}, \& {Tian}}]{tian2020}
{Ren}, J.~J., {Raddi}, R., {Rebassa-Mansergas}, A., {et~al.} 2020, \apj, 905,
  38, \dodoi{10.3847/1538-4357/abc017}

\bibitem[{{Riello} {et~al.}(2021){Riello}, {De Angeli}, {Evans}, {Montegriffo},
  {Carrasco}, {Busso}, {Palaversa}, {Burgess}, {Diener}, {Davidson}, {Rowell},
  {Fabricius}, {Jordi}, {Bellazzini}, {Pancino}, {Harrison}, {Cacciari}, {van
  Leeuwen}, {Hambly}, {Hodgkin}, {Osborne}, {Altavilla}, {Barstow}, {Brown},
  {Castellani}, {Cowell}, {De Luise}, {Gilmore}, {Giuffrida}, {Hidalgo},
  {Holland}, {Marinoni}, {Pagani}, {Piersimoni}, {Pulone}, {Ragaini}, {Rainer},
  {Richards}, {Sanna}, {Walton}, {Weiler}, \& {Yoldas}}]{Riello2021}
{Riello}, M., {De Angeli}, F., {Evans}, D.~W., {et~al.} 2021, \aap, 649, A3,
  \dodoi{10.1051/0004-6361/202039587}

\bibitem[{{Ruiz-Lara} {et~al.}(2020){Ruiz-Lara}, {Gallart}, {Bernard}, \&
  {Cassisi}}]{Ruiz-Lara2020}
{Ruiz-Lara}, T., {Gallart}, C., {Bernard}, E.~J., \& {Cassisi}, S. 2020, Nature
  Astronomy, 4, 965, \dodoi{10.1038/s41550-020-1097-0}

\bibitem[{{Ruoyi} \& {Haibo}(2020)}]{Ruoyi2020}
{Ruoyi}, Z., \& {Haibo}, Y. 2020, \apjl, 905, L20,
  \dodoi{10.3847/2041-8213/abccc4}

\bibitem[{{Schlafly} \& {Finkbeiner}(2011)}]{Schlafly2011}
{Schlafly}, E.~F., \& {Finkbeiner}, D.~P. 2011, \apj, 737, 103,
  \dodoi{10.1088/0004-637X/737/2/103}

\bibitem[{{Schlegel} {et~al.}(1998){Schlegel}, {Finkbeiner}, \&
  {Davis}}]{SFD1998}
{Schlegel}, D.~J., {Finkbeiner}, D.~P., \& {Davis}, M. 1998, \apj, 500, 525,
  \dodoi{10.1086/305772}

\bibitem[{{Starkenburg} {et~al.}(2017){Starkenburg}, {Martin}, {Youakim},
  {Aguado}, {Allende Prieto}, {Arentsen}, {Bernard}, {Bonifacio}, {Caffau},
  {Carlberg}, {C{\^o}t{\'e}}, {Fouesneau}, {Fran{\c{c}}ois}, {Franke},
  {Gonz{\'a}lez Hern{\'a}ndez}, {Gwyn}, {Hill}, {Ibata}, {Jablonka},
  {Longeard}, {McConnachie}, {Navarro}, {S{\'a}nchez-Janssen}, {Tolstoy}, \&
  {Venn}}]{Starkenburg2017}
{Starkenburg}, E., {Martin}, N., {Youakim}, K., {et~al.} 2017, \mnras, 471,
  2587, \dodoi{10.1093/mnras/stx1068}

\bibitem[{{Steinmetz} {et~al.}(2006){Steinmetz}, {Zwitter}, {Siebert},
  {Watson}, {Freeman}, {Munari}, {Campbell}, {Williams}, {Seabroke}, {Wyse},
  {Parker}, {Bienaym{\'e}}, {Roeser}, {Gibson}, {Gilmore}, {Grebel}, {Helmi},
  {Navarro}, {Burton}, {Cass}, {Dawe}, {Fiegert}, {Hartley}, {Russell},
  {Saunders}, {Enke}, {Bailin}, {Binney}, {Bland-Hawthorn}, {Boeche}, {Dehnen},
  {Eisenstein}, {Evans}, {Fiorucci}, {Fulbright}, {Gerhard}, {Jauregi}, {Kelz},
  {Mijovi{\'c}}, {Minchev}, {Parmentier}, {Pe{\~n}arrubia}, {Quillen}, {Read},
  {Ruchti}, {Scholz}, {Siviero}, {Smith}, {Sordo}, {Veltz}, {Vidrih}, {von
  Berlepsch}, {Boyle}, \& {Schilbach}}]{RAVE2006}
{Steinmetz}, M., {Zwitter}, T., {Siebert}, A., {et~al.} 2006, \aj, 132, 1645,
  \dodoi{10.1086/506564}

\bibitem[{{Weiler}(2018)}]{weiler2018}
{Weiler}, M. 2018, \aap, 617, A138, \dodoi{10.1051/0004-6361/201833462}

\bibitem[{{Whitten} {et~al.}(2019){Whitten}, {Placco}, {Beers}, {Chies-Santos},
  {Bonatto}, {Varela}, {Crist{\'o}bal-Hornillos}, {Ederoclite}, {Masseron},
  {Lee}, {Akras}, {Borges Fernandes}, {Caballero}, {Cenarro}, {Coelho},
  {Costa-Duarte}, {Daflon}, {Dupke}, {Lopes de Oliveira}, {L{\'o}pez-Sanjuan},
  {Mar{\'\i}n-Franch}, {Mendes de Oliveira}, {Moles}, {Orsi}, {Rossi},
  {Sodr{\'e}}, \& {V{\'a}zquez Rami{\'o}}}]{Whitten2019}
{Whitten}, D.~D., {Placco}, V.~M., {Beers}, T.~C., {et~al.} 2019, \aap, 622,
  A182, \dodoi{10.1051/0004-6361/201833368}

\bibitem[{{Whitten} {et~al.}(2021){Whitten}, {Placco}, {Beers}, {An}, {Lee},
  {Almeida-Fernandes}, {Herpich}, {Daflon}, {Barbosa}, {Perottoni}, {Rossi},
  {Tissera}, {Yoon}, {Youakim}, {Schoenell}, {Ribeiro}, \&
  {Kanaan}}]{Whitten2021}
---. 2021, \apj, 912, 147, \dodoi{10.3847/1538-4357/abee7e}

\bibitem[{{Wolf} {et~al.}(2018){Wolf}, {Onken}, {Luvaul}, {Schmidt}, {Bessell},
  {Chang}, {Da Costa}, {Mackey}, {Martin-Jones}, {Murphy}, {Preston}, {Scalzo},
  {Shao}, {Smillie}, {Tisserand}, {White}, \& {Yuan}}]{wolf2018}
{Wolf}, C., {Onken}, C.~A., {Luvaul}, L.~C., {et~al.} 2018, \pasa, 35, e010,
  \dodoi{10.1017/pasa.2018.5}

\bibitem[{{Wright} {et~al.}(2021){Wright}, {Lagos}, {Power}, \&
  {Correa}}]{Wright2021}
{Wright}, R.~J., {Lagos}, C. d.~P., {Power}, C., \& {Correa}, C.~A. 2021,
  \mnras, 504, 5702, \dodoi{10.1093/mnras/stab1057}

\bibitem[{{Wu} {et~al.}(2011){Wu}, {Luo}, {Li}, {Shi}, {Prugniel}, {Liang},
  {Zhao}, {Zhang}, {Bai}, {Wei}, {Dong}, {Zhang}, \& {Chen}}]{wu2011}
{Wu}, Y., {Luo}, A.~L., {Li}, H.-N., {et~al.} 2011, Research in Astronomy and
  Astrophysics, 11, 924, \dodoi{10.1088/1674-4527/11/8/006}

\bibitem[{{Xiang} {et~al.}(2019){Xiang}, {Ting}, {Rix}, {Sandford}, {Buder},
  {Lind}, {Liu}, {Shi}, \& {Zhang}}]{Xiang2019}
{Xiang}, M., {Ting}, Y.-S., {Rix}, H.-W., {et~al.} 2019, \apjs, 245, 34,
  \dodoi{10.3847/1538-4365/ab5364}

\bibitem[{{Yang} {et~al.}(2021){Yang}, {Yuan}, {Zhang}, {Niu}, {Huang}, {Duan},
  \& {Fang}}]{Yang2021}
{Yang}, L., {Yuan}, H., {Zhang}, R., {et~al.} 2021, \apjl, 908, L24,
  \dodoi{10.3847/2041-8213/abdbae}

\bibitem[{{Yanny} {et~al.}(2009){Yanny}, {Rockosi}, {Newberg}, {Knapp},
  {Adelman-McCarthy}, {Alcorn}, {Allam}, {Allende Prieto}, {An}, {Anderson},
  {Anderson}, {Bailer-Jones}, {Bastian}, {Beers}, {Bell}, {Belokurov},
  {Bizyaev}, {Blythe}, {Bochanski}, {Boroski}, {Brinchmann}, {Brinkmann},
  {Brewington}, {Carey}, {Cudworth}, {Evans}, {Evans}, {Gates}, {G{\"a}nsicke},
  {Gillespie}, {Gilmore}, {Nebot Gomez-Moran}, {Grebel}, {Greenwell}, {Gunn},
  {Jordan}, {Jordan}, {Harding}, {Harris}, {Hendry}, {Holder}, {Ivans},
  {Ivezi{\v{c}}}, {Jester}, {Johnson}, {Kent}, {Kleinman}, {Kniazev},
  {Krzesinski}, {Kron}, {Kuropatkin}, {Lebedeva}, {Lee}, {French Leger},
  {L{\'e}pine}, {Levine}, {Lin}, {Long}, {Loomis}, {Lupton}, {Malanushenko},
  {Malanushenko}, {Margon}, {Martinez-Delgado}, {McGehee}, {Monet}, {Morrison},
  {Munn}, {Neilsen}, {Nitta}, {Norris}, {Oravetz}, {Owen}, {Padmanabhan},
  {Pan}, {Peterson}, {Pier}, {Platson}, {Re Fiorentin}, {Richards}, {Rix},
  {Schlegel}, {Schneider}, {Schreiber}, {Schwope}, {Sibley}, {Simmons},
  {Snedden}, {Allyn Smith}, {Stark}, {Stauffer}, {Steinmetz}, {Stoughton},
  {SubbaRao}, {Szalay}, {Szkody}, {Thakar}, {Sivarani}, {Tucker}, {Uomoto},
  {Vanden Berk}, {Vidrih}, {Wadadekar}, {Watters}, {Wilhelm}, {Wyse}, {Yarger},
  \& {Zucker}}]{Yanny2009}
{Yanny}, B., {Rockosi}, C., {Newberg}, H.~J., {et~al.} 2009, \aj, 137, 4377,
  \dodoi{10.1088/0004-6256/137/5/4377}

\bibitem[{York {et~al.}(2000)York, Adelman, John E.~Anderson, Anderson, Annis,
  Bahcall, Bakken, Barkhouser, Bastian, Berman, Boroski, Bracker, Briegel,
  Briggs, Brinkmann, Brunner, Burles, Carey, Carr, Castander, Chen, Colestock,
  Connolly, Crocker, Csabai, Czarapata, Davis, Doi, Dombeck, Eisenstein,
  Ellman, Elms, Evans, Fan, Federwitz, Fiscelli, Friedman, Frieman, Fukugita,
  Gillespie, Gunn, Gurbani, de~Haas, Haldeman, Harris, Hayes, Heckman,
  Hennessy, Hindsley, Holm, Holmgren, hao Huang, Hull, Husby, Ichikawa,
  Ichikawa, Ivezi{\'{c}}, Kent, Kim, Kinney, Klaene, Kleinman, Kleinman, Knapp,
  Korienek, Kron, Kunszt, Lamb, Lee, Leger, Limmongkol, Lindenmeyer, Long,
  Loomis, Loveday, Lucinio, Lupton, MacKinnon, Mannery, Mantsch, Margon,
  McGehee, McKay, Meiksin, Merelli, Monet, Munn, Narayanan, Nash, Neilsen,
  Neswold, Newberg, Nichol, Nicinski, Nonino, Okada, Okamura, Ostriker, Owen,
  Pauls, Peoples, Peterson, Petravick, Pier, Pope, Pordes, Prosapio,
  Rechenmacher, Quinn, Richards, Richmond, Rivetta, Rockosi, Ruthmansdorfer,
  Sandford, Schlegel, Schneider, Sekiguchi, Sergey, Shimasaku, Siegmund, Smee,
  Smith, Snedden, Stone, Stoughton, Strauss, Stubbs, SubbaRao, Szalay, Szapudi,
  Szokoly, Thakar, Tremonti, Tucker, Uomoto, Berk, Vogeley, Waddell, i~Wang,
  Watanabe, Weinberg, Yanny, \& Yasuda}]{SDSS2000}
York, D.~G., Adelman, J., John E.~Anderson, J., {et~al.} 2000, The Astronomical
  Journal, 120, 1579, \dodoi{10.1086/301513}

\bibitem[{{Yuan} {et~al.}(2015{\natexlab{a}}){Yuan}, {Liu}, {Xiang}, {Huang},
  \& {Chen}}]{yuan2015metal}
{Yuan}, H., {Liu}, X., {Xiang}, M., {Huang}, Y., \& {Chen}, B.
  2015{\natexlab{a}}, \apj, 799, 134, \dodoi{10.1088/0004-637X/799/2/134}

\bibitem[{{Yuan} {et~al.}(2015{\natexlab{b}}){Yuan}, {Liu}, {Xiang}, {Huang},
  \& {Chen}}]{yuan2015FGK}
---. 2015{\natexlab{b}}, \apj, 803, 13, \dodoi{10.1088/0004-637X/803/1/13}

\bibitem[{{Yuan} {et~al.}(2015{\natexlab{c}}){Yuan}, {Liu}, {Xiang}, {Huang},
  {Chen}, {Wu}, {Hou}, \& {Zhang}}]{yuan2015binary}
{Yuan}, H., {Liu}, X., {Xiang}, M., {et~al.} 2015{\natexlab{c}}, \apj, 799,
  135, \dodoi{10.1088/0004-637X/799/2/135}

\bibitem[{{Yuan} {et~al.}(2015{\natexlab{d}}){Yuan}, {Liu}, {Xiang}, {Huang},
  {Zhang}, \& {Chen}}]{yuan2015recalibrated}
---. 2015{\natexlab{d}}, \apj, 799, 133, \dodoi{10.1088/0004-637X/799/2/133}

\bibitem[{{Yuan} {et~al.}(2013){Yuan}, {Liu}, \& {Xiang}}]{yuan2013}
{Yuan}, H.~B., {Liu}, X.~W., \& {Xiang}, M.~S. 2013, \mnras, 430, 2188,
  \dodoi{10.1093/mnras/stt039}

\bibitem[{{Zhang} {et~al.}(2021){Zhang}, {Yuan}, {Liu}, {Xiang}, {Huang}, \&
  {Chen}}]{zhang2021redgiant}
{Zhang}, R., {Yuan}, H., {Liu}, X., {et~al.} 2021, arXiv e-prints,
  arXiv:2109.06390.
\newblock \doarXiv{2109.06390}

\end{thebibliography}

\end{document}